\documentclass{aa}
\usepackage{graphicx}
\usepackage{natbib}
\usepackage{rotating}
\usepackage{bbm}

\bibpunct{(}{)}{;}{a}{}{,}
\newcommand{\be}{\begin{equation}}
\newcommand{\ee}{\end{equation}}
\newcommand{\vek}[1]{\mbox{\boldmath$#1$}}
\newcommand{\mytheta}{{\mbox{\boldmath$\vartheta$}}}
\newcommand{\myarcsec}{\hbox{$.\!\!^{\prime\prime}$}}
\newcommand{\myarcmin}{\hbox{$.\!\!^{\prime}$}}
\newcommand{\dd}{\textrm{d}}
\def\sun{\hbox{$\odot$}}
\begin{document}
   \title{Searching for galaxy clusters using the aperture mass statistics in 50 VLT fields\thanks{based on observations with FORS1@VLT operated by ESO (programme 63.O-0039A)}}

   \author{M. Hetterscheidt\inst{1}
          \and
          T. Erben\inst{1} \and P. Schneider\inst{1} \and R. Maoli\inst{2,4} \and L. Van Waerbeke\inst{3} \and Y. Mellier\inst{4,5}
          }

   \offprints{Marco Hetterscheidt, 
e-mail: mhetter@astro.uni-bonn.de}

   \institute{$^{1}$Institut f\"ur Astrophysik und Extraterrestrische Forschung (IAEF), Universit\"at Bonn, Auf dem H\"ugel 71, D-53121 Bonn, Germany\\
              $^{2}$Department of Physics, University ``La Sapienza", P.le A. Moro 2, 00185, Rome, Italy\\
              $^{3}$Department of Physics and Astronomy, University of British Columbia, Agricultural Road 6224, Vancouver, V6T 1Z1, B.C., CANADA\\ 
              $^{4}$Institut d'Astrophysique de Paris, UMR7095 CNRS, Universit\'e Pierre \& Marie Curie, 98 bis boulevard Arago, 75014 Paris, France\\
              $^{5}$Observatoire de Paris. LERMA. 61, avenue de l'Observatoire, 75014 Paris, France.}

   \date{Received 29 April 2005/ Accepted 30 June 2005}

   \abstract{
     Application of the aperture mass ($M_{\rm ap}$-) statistics provides 
     a weak lensing method for the detection of cluster-sized dark matter halos.
     We present a new aperture filter function and
     maximise the effectiveness of the $M_{\rm ap}$-statistics to detect cluster-sized halos
     using analytical models.
     We then use weak lensing mock catalogues generated from 
     ray-tracing through $N$-body simulations, to analyse the effect of image treatment on
     the expected number density of halos.
     Using the $M_{\rm ap}$-statistics, the aperture radius is typically several arcminutes, hence
     the aperture often lies partly outside a data field, 
     consequently the signal-to-noise ratio of a halo detection decreases.
     We study these border effects analytically and by using mock catalogues.
     We find that the expected number density of halos decreases by a factor of two
     if the size of a field is comparable to the diameter of the aperture used.
     We finally report on the results of a weak lensing cluster search applying the $M_{\rm ap}$-statistics 
     to 50 randomly selected fields which were observed with FORS1 at the VLT.
     Altogether the 50 VLT fields cover an area of 0.64 square degrees. 
     The $I$-band images were taken under excellent seeing conditions (average seeing $\approx 0\myarcsec 6$)
     which results in a high number density of galaxies used for the weak lensing 
     analysis ($n\approx 26\,{\rm arcmin}^{-2}$).
     In five of the VLT fields, we detect a significant $M_{\rm ap}$-signal which coincides with an 
     overdensity of the light distribution.
     These detections are thus excellent candidates for shear-selected clusters.
   }

   \maketitle

\section{Introduction}

Galaxy clusters are the largest collapsed structures in the Universe and formed due to the amplification of primordial density inhomogeneities and subsequent merging processes. 
Measuring their distribution and structures over a large redshift interval provides crucial information about the history and the large-scale structure of the Universe.
Because their formation and evolution are almost exclusively driven by gravity, the number density, main cluster properties and their dependence on redshift can be predicted analytically and by numerical simulations \citep{prs74,lac93,nfw96,nfw97,jfw01}.
There are several observational methods to test the theoretical predictions. 
The traditional way of obtaining the number density and the main properties of galaxy clusters is using direct observable quantities, like luminosity, temperature of the intra-cluster gas and the line-of-sight velocity dispersion of the cluster members.
The disadvantage of these methods is that they are based on simplified assumptions, such as hydrostatic equilibrium of the intra-cluster gas, virial equilibrium and/or spherical symmetry.
       
Weak gravitational lensing provides an opportunity to measure the (projected) mass distribution without making any of the assumptions mentioned above \citep{kai94,sch96_2}.
Furthermore, it is totally independent of the baryonic content. 
Aside from the analysis of already known mass concentrations like galaxy clusters, weak lensing techniques can be used to perform a blind search for hitherto unknown mass concentrations, with which it may then be possible to compile a purely shear-selected cluster sample.
Due to high demands on data quality, only in recent years several groups have started to use this method.
So far, only a few galaxy cluster candidates have been reported in the literature \citep{mhs02,wtm01,wtm03,ses03,ses04,dpl03,dsc04}. 
These candidates could be identified with overdensities of bright galaxies showing the presence of `regular' clusters.
Moreover, four of them are spectroscopically confirmed.
In addition, three shear-selected mass concentrations not associated with an optical counterpart, have been reported \citep{ewm00,umf00,dpl03}. 
Further investigations are necessary to confirm or discard these possible cases of mass concentrations with an unusually high mass-to-light ratio, 
given that even one of them would have profound impact on our understanding of the evolution of dark matter halos and their baryonic content, see \citet{ves05}.

Assuming a random distribution of galaxy orientations in the case of no lensing, a coherent alignment of galaxy ellipticities could indicate a mass concentration.  
A quantitative way to measure this alignment is the so-called aperture mass ($M_{\rm ap}$) statistics \citep{sch96_2}. 
In this paper we analyse the ability of the $M_{\rm ap}$-statistics to detect massive mass concentrations.

\citet{krs99} performed calculations of the expected number density of halos using $M_{\rm ap}$ with the polynomial filter function introduced by \citet{svj98}.
In this work we use a more effective filter function, which has already been applied to observational data by \citet{sch04} and \citet{ses04}.
We calibrate the filter function to detect a maximum number density of cluster-sized dark matter halos assuming an universal density profile.
We then use this filter function and apply the $M_{\rm ap}$-statistics to simulations and to a data set obtained with the VLT.

\citet{wvm02} and \citet{hty04} used numerical simulations to determine the expected number density of halos and examined completeness and efficiency in a weak lensing survey taking into account the noise caused by the ellipticity dispersion of background galaxies and the projection effects by large-scale structure.
\citet{hes05} introduced the so-called `Tomographic Matched Filtering' scheme which combines tomography using redshift information of background galaxies and matched filtering.
In their work it is shown that with photometric redshift information at hand it is possible to enhance the number density of clusters with high signal-to-noise ratio significantly. 
In the paper by \citet{mmb04} a nice derivation of a filter function is given with the aim of separating dark matter halos from spurious peaks in weak lensing maps caused by large-scale structure lensing.
They use numerical simulations to show qualitatively the sensitivity and reliability of this new filter function and compare the results with the conventionally used polynomial filter function.

In the present paper we describe the creation of synthetic images from ray-tracing through $N$-body simulations.
In addition to previous work we use these synthetic images to analyse how border effects, image treatment (galaxy detection and their ellipticity determination, PSF correction and shear estimation) and weighting affect the signal-to-noise ratio of peaks and their expected number density in the weak lensing maps. 
We finally report on the results of a cluster search applying the $M_{\rm ap}$-statistics to 50 VLT fields and compare these to the simulations. 
Criteria are presented with which peaks resulting from real clusters can possibly be distinguished from noise peaks in the weak lensing maps.

The paper is organised as follows.
In Sect. \ref{sec:detect} we describe the $M_{\rm ap}$-statistics and the new filter function, and calculate the detectability and number density of halos.
Numerical simulations are used in Sect. \ref{sec:sim} to create synthetic images in order to determine the expected number density of halos and to study the effect of image treatment on the signal-to-noise ratio of peaks in the weak lensing maps and the resulting change in the expected number density of halos. 
In Sect. \ref{sec:bordereffect} we present the creation of 300 VLT-sized synthetic images with which the border effects are studied.
The observed VLT fields are analysed in Sect. \ref{sec:VLT} and the number density obtained is compared with expectations determined in Sect. \ref{sec:bordereffect}.
In Sect. \ref{sec:candidates} all cluster candidates are presented and studied in detail.
A summary and conclusions are given in Sect. \ref{sec:conclusion}.
Appendix \ref{appendixa} provides the reader with all the formulas we used to calculate the number density of halos.

%
%
\section{Using the aperture mass for cluster detection}
\label{sec:detect}
In the following we use standard lensing notation. 
For a broader introduction to the topic, see for example \citet{bas01}.
\subsection{Introduction}
\label{sec:intro}
The gravitational field of a cluster-sized mass concentration causes a distortion of the background galaxy images, which is  revealed as a tangential alignment with respect to the centre of the mass concentration.
We use the aperture mass statistics, introduced by \cite{sch96_2}, to quantify the detectability of cluster-sized dark matter halos.
The aperture mass, $M_{\rm ap}$, is defined as the spatially filtered projected mass distribution, $\kappa$, inside a circular aperture of angular radius $\theta_0$ at a position $\vek{\xi}$,
\be 
M_{\rm ap}(\vek{\xi})\equiv \int\textrm{d}^2\theta\,\kappa(\vek{\theta})\,U(|\vek{\theta}-\vek{\xi}|),
\ee 
where $U$ is a radially symmetric continuous weight function.
Using a compensated filter function of radius $\theta_0$,
\be
\int_0^{\theta_0} \dd \theta \,\theta\, U(\theta)=0,
\label{compensate}
\ee 
one can express $M_{\rm ap}$ in terms of the tangential shear $\gamma_{\rm t}$
\be
   M_{\rm ap}(\vek{\xi})=\int\textrm{d}^2\theta\,\gamma_{\rm t}(\vek{\theta};\vek{\xi})\,Q(|\vek{\theta}-\vek{\xi}|).
\label{mapgammat}
\ee
The observable quantity $\gamma_{\rm t}(\vek{\theta};\vek{\xi})=-{\rm Re}[\gamma(\vek{\theta})e^{-2{\rm i}\phi}]$ is the tangential component of the shear at a position $\vek{\theta}-\vek{\xi}=|\vek{\theta}-\vek{\xi}|(\cos\phi,\sin\phi)$, where $\phi$ is the polar angle of $\vek{\theta}-\vek{\xi}$. 
The filter functions $Q$ and $U$ are related through
\be
Q(\vartheta)=\frac{2}{\vartheta^2}\int^\vartheta_0\dd\vartheta^\prime\, \vartheta^\prime\, U(\vartheta^\prime)-U(\vartheta).
\label{relation}
\ee
%
\cite{sch96_2} showed that the variance $\sigma_{\rm c}$ of $M_{\rm ap}$ is computable analytically as
\be
\sigma^2_{\rm c}(\theta_0)=\frac{\pi \sigma^2_\epsilon}{n}\int_0^{\theta_0} \dd\theta\, \theta\, Q^2(\theta),
\label{sigmac1}
\ee
where $\sigma_\epsilon$ is the ellipticity dispersion of galaxies and $n$ is the number density of galaxies in the considered aperture.
%
The signal-to-noise ratio ($snr$) is then
\be
snr=\frac{M_{\rm ap}}{\sigma_{\rm c}}=\sqrt{\frac{n}{\pi\,\sigma_\epsilon^2}}\,\frac{\int\textrm{d}^2\theta\,\gamma_{\rm t}(\theta)\,Q(\theta)}
     {\sqrt{\int_0^{\theta_0} \dd\theta\, \theta\, Q^2(\theta)}}.
\label{snrmap}
\ee
The main advantages of using the aperture mass statistics to search for clusters are that $M_{\rm ap}$ can be derived from the shear in a finite region and $M_{\rm ap}$ is not influenced by the mass-sheet degeneracy (both points follow from the fact that the filter function $U$ is compensated).
In addition, the application of $M_{\rm ap}$ to observational data is straightforward (one has to change the integral in Eq. (\ref{mapgammat}) into a sum over galaxy images) and the error analysis is simple, see Eq. (\ref{snrmap}).    
%
%
\subsection{An adapted filter}
\label{sec:optimal}
In \cite{svj98} a family of polynomial filter functions which fulfil the conditions (\ref{compensate}) and (\ref{relation}) were introduced for mathematical convenience,
\be
U(\theta;\theta_0)=\frac{(l+2)^2}{\pi\,\theta_0^2}\,\left(1-(\theta/\theta_0)^2\right)^l \left(\frac{1}{l+2}-(\theta/\theta_0)^2\right),
\ee
which corresponds to 
\be
 Q(\theta;\theta_0)=\frac{(1+l)(2+l)}{\pi\,\theta_0^2}\,(\theta/\theta_0)^2\, \left(1-(\theta/\theta_0)^2\right)^2,
\label{filter}
\ee
where $\theta$ is the projected angular distance on the sky from the aperture centre, $\theta_0$ is the filter radius.
Throughout the paper we will choose $l=1$.
However, in order to find the maximum number of dark matter halos one should use a filter function $Q$ which maximizes the $snr$.
According to the Cauchy-Schwarz inequality, the optimal choice for radially symmetric halos is $Q(\theta)\propto\gamma_{\rm t}(\theta)$, see \citet{sch96_2}.

Assuming the universal density profile found by Navarro, Frenk \& White (\citet{nfw96,nfw97}, NFW-profile in the following), a reasonable choice for the filter function $Q$ for observational purposes has been introduced by \cite{sch04},
\be
Q(x)=\left(1+{\rm e}^{a-bx}+{\rm e}^{-c+dx}\right)^{-1}\,\frac{{\rm tanh}(x/x_{\rm c})}{\pi \theta_0^2 (x/x_{\rm c})},
\label{Mfilter}
\ee
with $x:=\theta/\theta_0$.
The filter function, which we refer to as `halo-filter' in the following, approximately follows the tangential shear profile of an NFW-halo over a large $x$-range and is mathematically simple.
We choose the values $a=6$ and $b=150$ so that $Q$ exponentially drops to zero at $x=0$.
Furthermore, a choice of e.g. $c=47$ and $d=50$ results in an exponential cut-off around $x=1$.
The filter function has the nice properties that it downweights the inner part of a cluster which is often associated with bright galaxies (so no faint background galaxies are visible in the centre) and that smooth weak lensing maps are obtained.

\begin{figure}
\centering
\resizebox{\hsize}{!}{\includegraphics[width=\textwidth,clip]{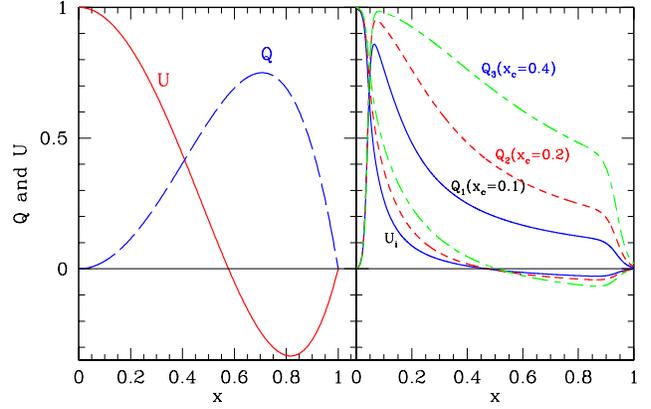}}
\caption{Left panel: the polynomial filter. 
Right panel: halo-filter for different parameters $x_{\rm c}$ with exponential cut off at both ends. 
The quantity $x$ is the normalized filter radius.
The maximum of all U-filters is normalized to 1.}
\label{fig:UundQ}
\end{figure}

The parameter $x_{\rm c}$ changes the shape of the filter in such a way that more weight is placed at smaller radii for smaller values of $x_{\rm c}$.
In the right panel of Fig. \ref{fig:UundQ} the filter function $Q$ and the corresponding filter $U$ are shown for different values of $x_{\rm c}$. 
In the left panel of this figure the polynomial filter functions introduced by \cite{svj98} are displayed for comparison.
%
%
%

In order to obtain predictions for the expected signal-to-noise ratios
for halos of mass $M$ at redshift $z_{\rm l}$, we have to
specify a mass model $\kappa(\mytheta, z_{\rm l}, z_{\rm s})$ and a
distribution for the source redshifts $p(z_{\rm s})$. 
The expected $M_{\rm ap}$-signal is then given by
\be
   M_{\rm ap}=\int {\rm d}^2\mytheta\int_{z_{\rm l}}^{\infty} \dd z_{\rm s}\,\kappa(\mytheta, z_{\rm l}, z_{\rm s})\,U(\vartheta)\,p(z_{\rm s}).
\label{zmap}
\ee
The noise is simply given by $\sigma_{\rm c}$ in (\ref{sigmac1}).
To calculate the mass profile we closely follow the work of \citet{hty04}.
We consider a truncated NFW-profile, see \citet{taj03}, and use the concentration parameter introduced by \citet{bks01}.
In Appendix A all equations to calculate the surface mass density $\kappa$ are listed. 

\begin{figure}
\centering
\resizebox{\hsize}{!}{\includegraphics[width=\textwidth,clip]{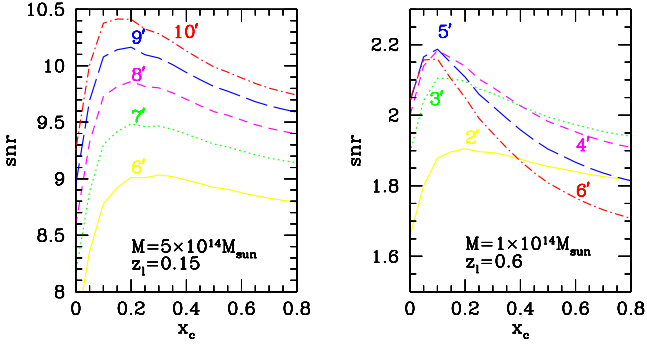}}
\caption{The dependence of the $snr$ on the filter parameter $x_{\rm c}$ for different filter radii $\theta_0$ for two different cluster masses and redshifts. 
Left panel: high cluster mass, low redshift. 
Right panel: low cluster mass, high redshift.}
\label{fig:xc}
\end{figure}

\begin{figure}
\resizebox{\hsize}{!}{
\setlength{\unitlength}{0.25cm}
\begin{picture}(36,36)
\put(1,1){\includegraphics[width=35\unitlength]{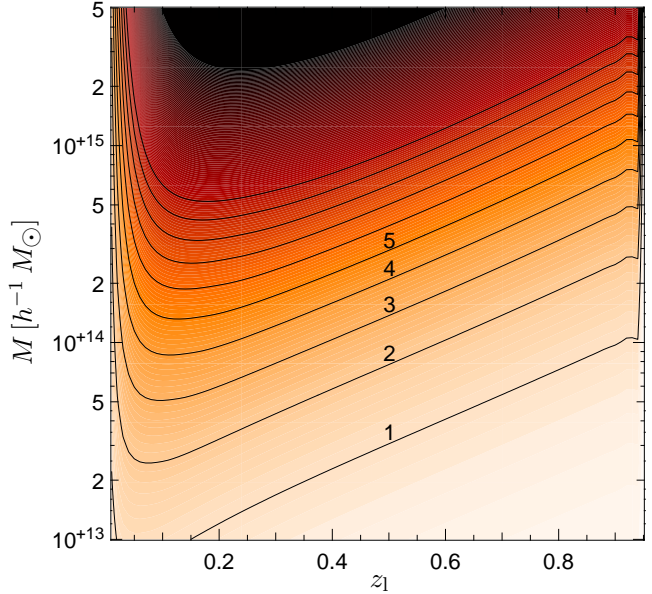}}
\put(21,1){\makebox(0,0)[t]{\large $z_{\rm l}$}}
\put(1,17){\makebox(0,0){\large \begin{sideways} $M\, [h^{-1}\,M_{\sun}]$ \end{sideways}}}
\end{picture}
}
\caption{Detectability of mass concentrations with an NFW-profile for different masses $M$ and redshifts $z_{\rm l}$. The contour lines indicate the $snr$. 
We assume $\Omega_\Lambda=0.7$, $\Omega_{\rm m}=0.3$, number density of galaxies $n=30/{\rm arcmin}^2$ and an ellipticity dispersion of $\sigma_\epsilon=0.4$.
The filter radius is chosen to be $\theta_0=6^\prime$ and $x_{\rm c}=0.15$.}
\label{fig:snrMatrix}
\end{figure}

To use the filter function (\ref{Mfilter}) we calculate how the parameter $x_{\rm c}$  changes the $snr$ for different halo masses, redshifts and filter radii; see Fig. \ref{fig:xc} for an example.
It turns out that a good choice for all reasonable combinations of halo masses, redshifts and filter radii is $x_{\rm c}=0.15$. 
Fixing $x_{\rm c}$, and using the filter radius $\theta_0=6^\prime$ (this filter radius is the optimal choice to find the maximum number of halos in the redshift range $z\in [0;0.95]$, see Fig. \ref{fig:MaxNumberR}) we calculate the $snr$ for different masses and redshifts of halos, given the redshift distribution of background galaxies introduced by \citet{bbs96},
\be
p(z)=\frac{\beta}{\Gamma[(1+\alpha)/\beta]\,z_0}\left(\frac{z}{z_0}\right)^\alpha \exp\left[-\left(\frac{z}{z_0}\right)^\beta\right],
\label{redshiftdistribution}
\ee
where $\Gamma$ is the Gamma-function, and $\alpha=2$, $\beta=1.5$ and $z_0=0.8$, which results in a mean redshift $\langle z\rangle=1.2$ \citep{wkl01}.
Furthermore, we assume in the following a typical weak lensing survey with a number density of galaxies of $n=30\,{\rm arcmin}^{-2}$ and an ellipticity dispersion of $\sigma_\epsilon=0.4$.
The results are shown in Fig. \ref{fig:snrMatrix}.
With the halo-filter function we should be able to detect halos with a signal-to-noise ratio larger than 4 down to masses of $10^{14}M_{\sun}$ for redshifts lower than 0.3.  
%
%
%
%
%
%
%
%
\subsection{Number density of halos}
\label{sec:number}
\begin{figure}
\centering
\resizebox{\hsize}{!}{\includegraphics[width=\textwidth,clip]{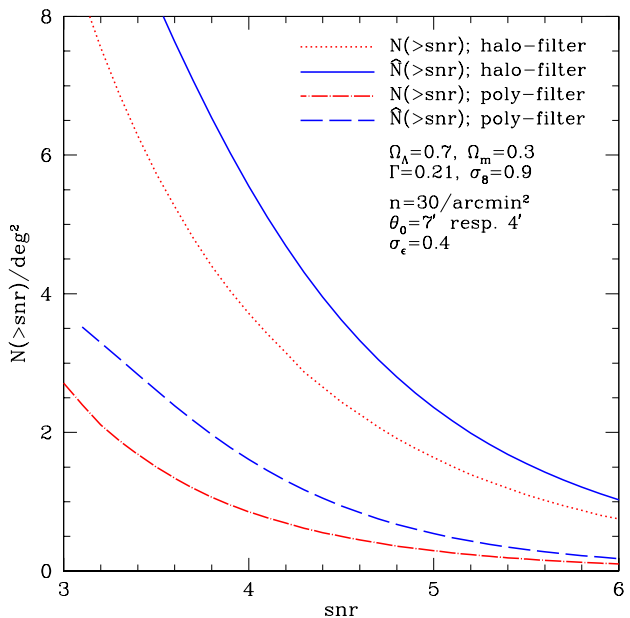}}
\caption{Theoretical number, $N$, and the observable number, $\hat N$, of halos per square degree with a signal-to-noise ratio larger than $snr$, calculated for the halo-filter ($\theta_0=6^\prime$) and, as a comparison, for the polynomial filter ($\theta_0=4^\prime$).
The observable number density is obtained by convolving the theoretical number density with the Gaussian distribution $p(\Delta M_{\rm ap};\theta_0)$ (\ref{gauss}).}
\label{fig:numberHalo}
\end{figure}

\begin{figure}
\centering
\resizebox{\hsize}{!}{\includegraphics[width=\textwidth,clip]{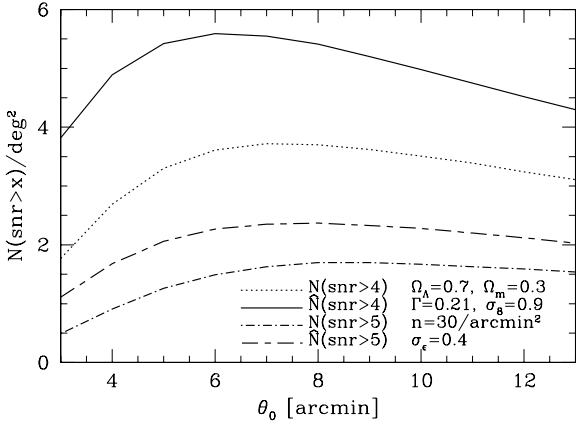}}
\caption{Dependence of the number of halos per square degree with a signal-to-noise ratio larger than 4 and larger than 5, respectively, on the aperture radius of the halo-filter.
The optimal choice for the halo filter radius to find the maximal number of halos with a $snr$ larger than 4 is $\theta_0=6^\prime$. 
}
\label{fig:MaxNumberR}
\end{figure}

To calculate the number density of significant peaks in the aperture mass map above a given $snr$-threshold, $N(>snr_{\rm t})$, resulting from real halos, we closely follow the work of \citet{krs99}.
It is assumed that dark matter halos are distributed according to the Press-Schechter theory.
We use the fitting formulae given in \citet{nfw97} to compute the number density $N$ of halos.
The aperture mass is a monotonically increasing function of halo mass $M$ for constant values of the lens redshift $z_{\rm l}$ and filter radius $\theta_0$, therefore it can be inverted for a given threshold value $M_{\rm ap}^{\rm t}$.
As the noise in Eq. (\ref{sigmac1}) only depends on the filter function in use, the $snr$-threshold value is $snr_{\rm t}=M_{\rm ap}^{\rm t}/\sigma_{\rm c}$. 
The number of halos in a given proper volume with a mass greater than $M_{\rm t}(snr_{\rm t},z_{\rm l},\theta_0)$, and thus a $snr$ greater than $snr_{\rm t}$, is given by,
\be
N(>snr_{\rm t})=\int \dd V_{\rm p}(1+z_{\rm l})^3 \int_{M_{\rm t}}^\infty \dd M N_{\rm halo}(M,z_{\rm l})
\ee
\citep{krs99}.
The quantity $N_{\rm halo}\,\dd V_{\rm c}\dd M$ is the number of halos in the comoving volume $\dd V_{\rm c}$ with mass in the interval $\dd M$.
We now calculate the number density $N(>snr_{\rm t})$ of halos above a given $snr$-threshold, assuming the mass profile and observational parameters ($n=30/{\rm arcmin}^2$, $\sigma_\epsilon=0.4$, $\langle z\rangle=1.2$) of Sect. \ref{sec:optimal}.
As can be seen in Fig. \ref{fig:numberHalo}, the theoretical number density $N(>snr_{\rm t}=4)$ calculated for the halo-filter is $\approx 3$ times larger than for the polynomial filter.
In comparison to \citet{krs99} our results for $N(>snr_{\rm t})$ using the polynomial filter are much lower, because our assumed ellipticity dispersion is twice as large, our mean redshift is lower and we use a slightly different mass profile. 

Taking into account only the noise induced by the intrinsic ellipticity distribution of the background galaxies, we assume that the difference between the real value of $M_{\rm ap}$ and the measured $\hat M_{\rm ap}$ follows a Gaussian distribution,
\be
p(\Delta M_{\rm ap};\theta_0)=\frac{1}{\sqrt{2\pi}\sigma_{\rm c}}\,\exp\left[{-\frac{\Delta M_{\rm ap}^2}{2\,\sigma_{\rm c}^2}}\right].
\label{gauss}
\ee
The observable number density $\hat N$ of halos above a given $snr$-threshold is then obtained by convolving the theoretical number density $N$ by $p(\Delta M_{\rm ap})$, see \citet{krs99}.
This results in an increase in the number density of peaks of high significance, see Fig. \ref{fig:numberHalo} and Table \ref{tab:number1}.
This, however, is only an approximation.
The noise not only changes the peak height but also the peak position.
A neighbouring pixel of the original peak maximum (without noise) can be higher after adding noise,
so the given calculations are just an upper limit. 
A profound analysis of the influence of Gaussian random fields on the peak statistics can be found in \citet{vwa00} and \citet{jvw00} (however, for high $snr$ our approximation is sufficient). 
%
%
\begin{table}
\caption{The number of halos per 10 square degrees with a signal-to-noise ratio larger than 4 and 5 for the theoretical number density, $N$, the observable number density, $\hat N$ (both calculated with a filter radius of $\theta_0=6^\prime$) and the observable number density, for which it is assumed that the area is analysed with different filter radii, $\hat N_{\rm vari}$.}
\label{tab:number1}
\begin{center}
\begin{tabular}{c|c|c|c}
 $z_{\rm l}\in[0,0.95]$ & $N/10\,{\rm deg}^2$ & $\hat N/10\,{\rm deg}^2$ & $\hat N_{\rm vari}/10\,{\rm deg}^2$ \\
\hline
$snr>4$ & 37 & 55 & 61 \\
$snr>5$ & 16 & 23 & 25 \\
\end{tabular}
\end{center}
\end{table}
%
%
\begin{table}
\caption{The maximum number of halos with a signal-to-noise ratio larger than 4 per 10 square degrees per redshift interval. 
The value $\theta_0$ is the filter radius which maximises the number density for the given redshift interval.}
\label{tab:number2}
\begin{center}
\begin{tabular}{c|c|c}
 $z_{\rm l}$ & $\hat N_{\rm max}\,[10\,{\rm deg}^{-2}]$ & $\theta_0\,[{\rm arcmin}]$ \\
\hline
$[0.0,0.15]$ & 4.6 & 13 \\
$[0.15,0.25]$ & 14.7 & 8 \\
$[0.25,0.35]$ & 14.7 & 6 \\
$[0.35,0.45]$ & 12.3 & 6 \\
$[0.45,0.95]$ & 14.4 & 5 \\
\end{tabular}
\end{center}
\end{table}
%
%

\begin{figure}
\centering
\resizebox{\hsize}{!}{\includegraphics[width=\textwidth,clip]{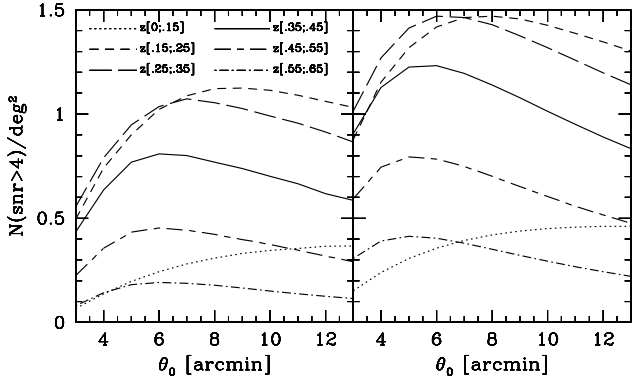}}
\caption{Dependence of the number of halos per square degree with a signal-to-noise ratio larger than 4 on the aperture radius for different redshift bins.
Left panel: theoretical number density ($N$). 
Right panel: observable number density ($\hat N$).}
\label{fig:scan}
\end{figure}

The values of projected angular radii on the sky of galaxy clusters having various virial radii and redshifts are very different.
Considering this fact one applies, in practice, the aperture mass statistics with varying filter radii.
To estimate the increase of the number density by applying different filter radii, we determine, for different redshift bins, the aperture radius for which we obtain the maximum number density of halos (Fig. \ref{fig:scan}). 
The maximum number is then added up (Table \ref{tab:number1}, \ref{tab:number2}).
We find that the expected number density of halos with a signal-to-noise ratio larger than four exceeds 61 per 10 square degree, which is only slightly larger than using a fixed filter radius of $\theta_0=6^\prime$ (55 per 10 square degree).
This is due to the fact that the radius for which one expects the maximal number of halos is approximately the same for every redshift interval (Fig. \ref{fig:scan}).

In the real world, not every galaxy cluster is relaxed and has an NFW-profile.
It is therefore difficult to estimate the difference in the number density between a fixed and variable filter radius because our filter function is optimised for NFW-profiles.
This is worth examining in detail with numerical simulations in a future paper.

In Table \ref{tab:number2} it can also be seen that even for redshifts larger than 0.45 we expect to find about 15 NFW-halos with a signal-to-noise ratio larger than 4 in a $10\,{\rm deg}^2$ survey.
%
%

%
%
%
%
\subsection{$M_{\rm ap}$ applied to images}
\label{sec:mapondata}
%
We now describe the application of the $M_{\rm ap}$-statistics to images.
It is straightforward to construct an
unbiased estimator $M'_{\rm ap}$ for the integral (\ref{mapgammat}) by a discrete sum
over observed galaxy ellipticities $\epsilon_{\rm t}$. Considering
the coordinate origin to be at the centre of the aperture this then reads as
\be
   M'_{\rm ap}=\frac{1}{n} \sum_i\epsilon_{\rm t}(\mytheta_i)Q(\vartheta_i),
\label{map}
\ee
where $n$ is the number density of galaxies in the considered aperture and $\epsilon_{\rm t}=-{\rm Re}[\epsilon(\mytheta)e^{-2{\rm i}\phi}]$ is the tangential ellipticity.
The discrete dispersion $\sigma_{\rm d}$ of $M_{\rm ap}$ in the case of no lensing can be calculated by squaring Eq. (\ref{map}) and taking the expectation value, which leads to
\be   
   \sigma_{\rm d}^2=\frac{\sigma_{\epsilon}^2}{2n^2}\sum_i Q^2(\vartheta_i),
\label{sigmad}
\ee
where $\sigma_{\epsilon}^2=\langle|\epsilon|^2\rangle$ is the ellipticity dispersion and $n$ the number density of galaxies.
We used the fact that the ellipticities of different images are not correlated ($\langle\epsilon_{{\rm t}i}\epsilon_{{\rm t}j}\rangle=\delta_{ij}\sigma_{\epsilon}^2/2$). 
In the case of weighting, the discrete aperture mass $M_{\rm ap}^\prime$ changes to
\be
  M^{\rm w}_{\rm ap}=\frac{\pi \theta_0^2 \sum_i\epsilon_{\rm t}(\mytheta_i)\,w_i\,Q(\vartheta_i)}{\sum_i w_i},
\label{mapweight}
\ee
and the discrete dispersion then reads
\be
   \sigma_{\rm d,w}^2=\frac{\pi^2 \theta_0^4 \sum_i|\epsilon(\mytheta_i)|^2\,w_i^2\,Q^2(\vartheta_i)}{2\,\left(\sum_i w_i\right)^2},
\ee
see \cite{sch04}.
The weighting factors $w_i$ are calculated from the uncertainty of $\epsilon_{\rm t}$ [see Eq. (\ref{gewichtung})].
A regular grid of aperture centres is now placed over a data field and the $snr$ is calculated for every grid point.
In this way weak lensing maps ($M_{\rm ap}$-maps) are obtained for every image in which mass concentrations are revealed as peaks in the map.

%
%
\section{Applying $M_{\rm ap}$ to numerical simulations}
\label{sec:sim}
In this section we investigate the ability of the aperture mass statistics to detect mass concentrations by applying $M_{\rm ap}$ to images obtained from ray-tracing through $\Lambda$CDM N-body simulations ($\Omega_\Lambda=0.7,\,\Omega_0=0.3,\,\sigma_8=0.9,\,h=0.7$).
These simulations were kindly made available by Takashi Hamana [details see \cite{hty04}].
We create twelve initial catalogues of randomly distributed galaxies using the programme \textit{stuff}\footnote{Available at:\\ \textup{http://terapix.iap.fr/cplt/oldSite/soft/stuff/}} (E. Bertin).
For a detailed description of the galaxy morphology and magnitude distribution see \cite{ewb01}.
The galaxies are assumed to be at a fixed redshift $z=1$ and are sheared according to the shear map of the ray-tracing simulations, meaning that we modify the intrinsic galaxy ellipticity $e$ by the shear $\gamma$ present at that position.
In the following these galaxy catalogues are called \textit{\textbf{input catalogues}}. 

The input catalogues are used to create synthetic images using the programme \textit{SkyMaker}\footnote{Available at:\\ \textup{http://terapix.iap.fr/cplt/oldSite/soft/skymaker/}} by E. Bertin.
A short description is given in \cite{ewb01}.
Twelve $30^\prime\times 30^\prime$ images resulting in a $3\,{\rm deg}^2$ survey are obtained for the twelve catalogues.  
These images are treated in exactly the same way as real data (like object detection, PSF correction, same cuts, weighting), see Sect. \ref{sec:catalogue}.
For the object detection we utilise the programme SExtractor \citep{bea96}.
An object is detected if 3 contiguous pixels are $1\sigma$ above the sky background (SExtractor parameter settings DETECT\_MINAREA=3, DETECT\_THRESH=1). 
The obtained galaxy catalogues are called \textit{\textbf{output catalogues}}.

To exclude the effect of false detections we only take into account those objects which are present in both catalogues. 
The mean ellipticity dispersion of the galaxies of the input and output catalogues is $\sigma_\epsilon=0.32$ and the mean galaxy number density is $n=19\,{\rm arcmin}^{-2}$.
Note that the galaxy number density quoted is the number density of background galaxies, as all galaxies are placed at redshift $z=1$.

In the following we apply the $M_{\rm ap}$-statistics using the halo-filter for two different filter radii ($\theta_0=1000\, {\rm pixels}=3\myarcmin 8$ and $\theta_0=1500\, {\rm pixels}=5\myarcmin 7$) on the input and output catalogues as described in Sect. \ref{sec:mapondata}, where the mesh size of the grid is $10^{\prime\prime}\times 10^{\prime\prime}$.
We examine various effects on the $snr$ of peaks and on the number density of significant peaks ($snr>3$) in the $M_{\rm ap}$-maps.
At this stage it is worth mentioning that significant peaks in $M_{\rm ap}$-maps can originate from
\begin{enumerate}
\item massive halos,
\item low mass halos which are boosted to high $snr$ by noise,
\item pure noise peaks,
\item projection effects (two or more halos at different redshift along almost the same line of sight are merged in the $M_{\rm ap}$-map due to the smoothing effect of the filter function),
\item real substructure in massive halos,
\item false high peaks in the vicinity of a massive cluster due to noise.
\end{enumerate} 
In this work we cannot distinguish between these cases.
Nevertheless, we can estimate the number of pure noise peaks by randomising the background galaxies in each field, see Sect. \ref{sec:noisepeaks}.
The influence of points 4-6 can be reduced if we consider only peaks with high signal-to-noise ratios ($snr>3$). 

For the analysis of various effects on the $snr$ of peaks in the $M_{\rm ap}$-maps we investigate the maps by eye.
For the statistics of the number density of halos we define a peak in the $M_{\rm ap}$-map as a pixel with maximum value compared to the surrounding 24 neighbour pixels.
By applying this definition and a mesh size of $10^{\prime\prime}\times 10^{\prime\prime}$ we reduce the number of false peaks in the vicinity of a massive cluster with $snr>3$ (point 6).
The difference between the number density of peaks found according to this definition and the investigation by eye is insignificant.

\subsection{Example of the simulation }
\begin{figure}
\centering
\resizebox{\hsize}{!}{\includegraphics[width=\textwidth]{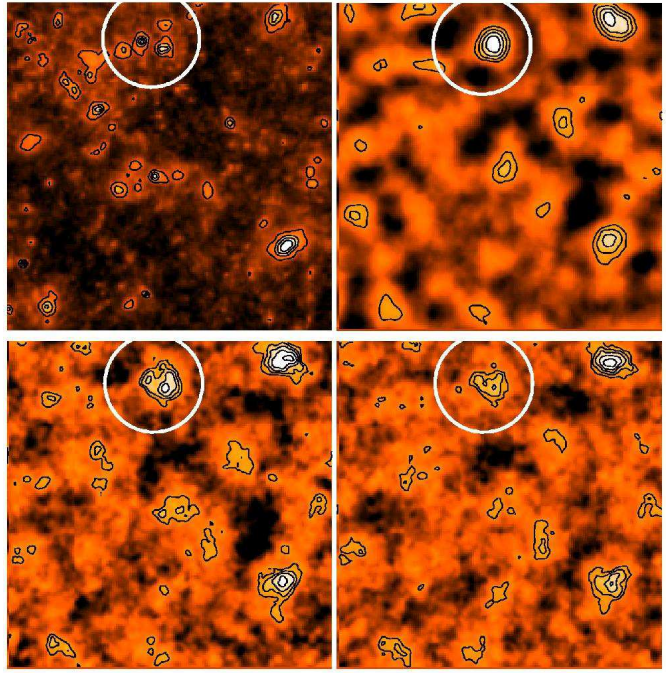}}
\caption{An example of the N-body simulations. \textbf{Upper left:} $\kappa$-map, \textbf{upper right:} $snr$-map, polynomial filter with $\theta_0=3\myarcmin 8$, \textbf{lower left:} $snr$-map, halo-filter with $\theta_0=5\myarcmin 7$, \textbf{lower right:} $snr$-map, halo-filter with $\theta_0=5\myarcmin 7$ after source extraction and PSF correction. 
Independent of the filter in use the most prominent features in the $\kappa$-map are detected with high significance.
White circle: in contrast to the polynomial filter the halo-filter resolves the three peaks in the $\kappa$-map. 
The size of each field is $30^\prime \times 30^\prime$.
Contour lines in kappa-map: 0.05, 0.1, 0.15 and 0.2; in all $snr$-maps: 2, 3, 4, 5, 6. 
 }
\label{fig:simulations}
\end{figure}

In Fig. \ref{fig:simulations} we present one of the $\kappa$-maps from the N-body simulations and compare it to the $snr$-maps for different filter functions and filter radii.
The most prominent features in the $\kappa$-map (upper left) are detected with a high significance in the $snr$-maps independent of the filter in use. 
The halo-filter not only detects mass concentrations with higher significance compared to the polynomial filter, but also resolves the maxima in the $\kappa$-map, indicated by the white circles in Fig. 7.   
%
%
\subsection{Two different halo-filter radii }
We apply the $M_{\rm ap}$-statistics for two different halo-filter radii ($\theta_0=3\myarcmin 8$ and $\theta_0=5\myarcmin 7$) to the input and output catalogues of the numerical simulations.
As expected from the calculations in the previous sections, the $snr$ on average increases for larger radius, see lower left diagram in Fig. \ref{fig:sim1} as example for the output catalogue.
With the larger radius more significant peaks are detected for the input catalogue.
This is illustrated in the upper right diagram of Fig. \ref{fig:sim2}.
11 peaks per square degree are detected with a $snr$ larger than four using the halo-filter with a radius of $\theta_0=5\myarcmin 7$, compared to only 6.5 using the halo-filter with a radius of $\theta_0=3\myarcmin 8$ (see also Tab. \ref{tab:numberdensity}).
%
%
\begin{table}
\caption{Number of peaks per one square degree with a signal-to-noise ratio larger than four. }
\label{tab:numberdensity}
\begin{center}
\begin{tabular}{l|r}
  & $N(snr>4)$  \\
\hline
halo-filter, input cat., $\theta_0=5\myarcmin 7$ & 11  \\
halo-filter, output cat., $\theta_0=5\myarcmin 7$ & 6  \\
halo-filter, output cat., weighting, $\theta_0=5\myarcmin 7$ & 7.5 \\
halo-filter, input cat., border effects, $\theta_0=5\myarcmin 7$ & 6 \\
halo-filter, output cat., border effects, $\theta_0=5\myarcmin 7$ & 3 \\
halo-filter, input cat. $\theta_0=3\myarcmin 8$ & 6.5 \\
halo-filter, input cat., border effects, $\theta_0=3\myarcmin 8$ & 4 \\
halo-filter, output cat., border effects, $\theta_0=3\myarcmin 8$ & 3 \\
poly-filter, input cat., $\theta_0=3\myarcmin 8$ & 2.5 \\
VLT images, $\theta_0=5\myarcmin 7$ & 2 \\
VLT images, $\theta_0=3\myarcmin 8$ & 3 \\
\end{tabular}
\end{center}
\end{table}
\subsection{Comparison between halo- \& polynomial filter and the analytical model}
\label{sec:comp}
The $snr$ of peaks in the $M_{\rm ap}$-maps obtained from the input catalogues is significantly lower if the polynomial filter is used, see lower right diagram in Fig. \ref{fig:sim1}.
As a consequence, the number density of peaks in the $M_{\rm ap}$-maps using the polynomial filter for a given $snr$ is significantly lower compared to the halo-filter.
We obtain 11 peaks per square degree with a $snr$ larger than four for the halo-filter ($\theta_0=5\myarcmin 7$), but only 2.5 for the polynomial filter ($\theta_0=3\myarcmin 8$), see lower right diagram in Fig. \ref{fig:sim2}.

In Fig. \ref{fig:sim2} the number densities of dark matter halos for the halo-filter and the polynomial filter obtained from the analytical model are also plotted.
In this section the same parameters are used for the calculations as for the simulations (fixed redshift $z=1$, $\theta_0=5\myarcmin 7$ for halo-filter, $\theta_0=3\myarcmin 8$ for polynomial filter, $\sigma_\epsilon=0.32$ and $n=19/{\rm arcmin}^2$).
Notable is the difference between the analytical model and the simulations.
For $snr$ less than $\approx 4$ the reasons are that peaks in the synthetic data can also originate from the effects listed in the introduction of Sect. \ref{sec:sim}.
The main reason for the difference for high $snr$ is that more massive clusters are present in the synthetic data (note that the area of the data only covers $3\,{\rm deg}^2$) as is indicated by the jump in the number density at $snr=5.6$ (halo-filter curve) which is then carried over to lower $snr$.
Another reason could be that we have used the Press-Schechter model to calculate the number density of halos, which underpredicts the more massive ones, see \cite{jfw01}.  

From the calculations we expect to detect 3 times more halos with the halo-filter than with the polynomial filter for a $snr > 4$.
But as the halo-filter resolves substructure (Fig. \ref{fig:simulations}; due to the fact that the halo-filter is much narrower than the polynomial filter) the number density of $M_{\rm ap}$-peaks obtained with the halo-filter is even 4.5 times larger than the number density obtained with the polynomial filter. 
\subsection{Noise peaks in weak lensing maps  }
\label{sec:noisepeaks}
\begin{figure}
\centering
\resizebox{\hsize}{!}{\includegraphics[width=\textwidth]{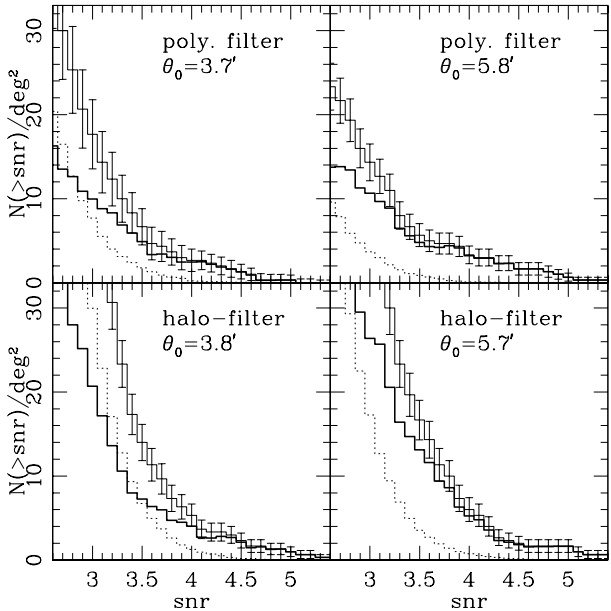}}
\caption{Comparison of number density of peaks in the $M_{\rm ap}$-maps per square degree with a signal-to-noise ratio larger than $snr$ of the output catalogue for halo- and polynomial filter and different filter radii.
Solid lines with error bars: total number of peaks per square degree in the $M_{\rm ap}$-maps. 
Dotted lines: number of noise peaks per square degree, resulting from randomisation of the orientation of background galaxies.
Bold solid lines: number of `real' peaks (difference between the total number of peaks in the $M_{\rm ap}$-maps and number due to randomisation) per square degree. 
The error bars are due to Poisson statistics obtained from 3 square degrees. 
For clarity they are only plotted for the total number density.
}
\label{fig:real_zufall}
\end{figure}

The number density of halos is contaminated by noise peaks caused by a chance alignment of background galaxies.
To quantify the number density of these noise peaks, we randomise the orientation of background galaxies, apply $M_{\rm ap}$ and repeat this procedure 20 times for each of the 12 output catalogues.
This is done for the two different filter and for two different filter scales, see Fig \ref{fig:real_zufall} and Tab. \ref{tab:numbernoise}.
As expected, the number density of noise peaks of a given $snr$-threshold is lower for a larger filter radius and is larger for the halo-filter compared to the polynomial filter, because it is narrower than the polynomial filter and therefore puts a high weight to a smaller number of galaxies.
\begin{table}
\caption{Number of noise peaks per one square degree.}
\label{tab:numbernoise}
\begin{center}
\begin{tabular}{l|r|r}
  & $N(>3)$ & $N(>4)$ \\
\hline
sim., halo-filter, $\theta_0=5\myarcmin 7$ & 17 & 0.75  \\
sim., halo-filter, $\theta_0=3\myarcmin 8$ & 30 & 1.28  \\
sim., halo-filter, border eff., $\theta_0=5\myarcmin 7$ & 21 & 0.81 \\
sim., poly. filter, $\theta_0=5\myarcmin 7$ & 3.7 & 0.10 \\
sim., poly. filter, $\theta_0=3\myarcmin 8$ & 7.7 & 0.22 \\
sim., poly. filter, border eff., $\theta_0=5\myarcmin 7$ & 3.4 & 0.10 \\
VLT images, halo-filter, $\theta_0=5\myarcmin 7$ & 20 & 0.68 \\
VLT images, halo-filter, $\theta_0=3\myarcmin 8$ & 30 & 0.85 \\
\end{tabular}
\end{center}
\end{table}

The number density of pure noise peaks is then subtracted from the total number density of peaks to estimate the number density of `real' peaks that are due to a real overdensity in the $\kappa$-map of the simulated field (Fig. \ref{fig:real_zufall}).

\begin{figure}
\centering
\resizebox{\hsize}{!}{\includegraphics[width=\textwidth]{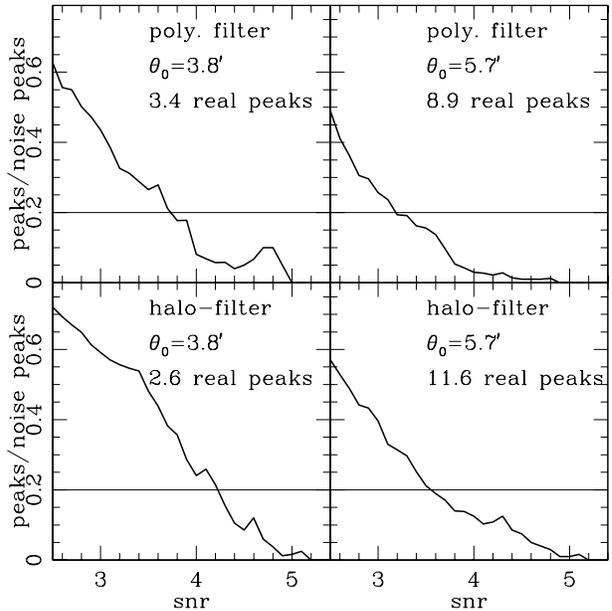}}
\caption{Contamination of noise peaks. 
Ratio between number of noise peaks and total number of peaks for the halo- and polynomial filter and for two different filter radii.  
The number of real peaks is the difference of the total number of peaks and the noise peaks at the $snr$ of 20\% noise peak contamination.
}
\label{fig:quo}
\end{figure}

We now answer the question of how many real peaks in the weak lensing maps we obtain if we only allow a contamination ratio (number of noise peaks to total number of peaks in the weak lensing maps) of 20\%. 
We plot this ratio in Fig. \ref{fig:quo}.
If we assume a contamination of 20\% we obtain 11.6 real peaks for the halo-filter and 8.9 real peaks for the polynomial filter.
If we take into account the contamination ratio, the difference of the efficiency of the two filter types is much less pronounced compared to the efficiency obtained if we only take into account a given $snr$-threshold, see last section.   
In other words, to compare the efficiency of different filters it is not sufficient to compare the number density of peaks for a given $snr$-threshold, but to compare the number density for a given contamination of noise peaks. 
%
%
%
%
%
%
\subsection{The effect of image treatment }
\label{sec:imagetreat}
Image treatment includes galaxy detection and the determination of galaxy quadrupole moments, PSF correction (anisotropy- and $P^{\rm g}$-correction) and catalogue filtering.
This will be described in detail in Sect. \ref{sec:catalogue}. 
In this work we discuss the effects of image treatment on the $snr$ and the change in the number density of $M_{\rm ap}$-peaks as a whole.
Detailed insights into the impact of different steps of the image treatment or KSB algorithm and its implementation on shear estimates will be presented for two different KSB-pipelines (Bonn and Edinburgh pipeline) in Hetterscheidt et al. (in prep.).

We apply the $M_{\rm ap}$-statistics to the input and output catalogues and compare the change of the $snr$ of peaks in the $M_{\rm ap}$-maps, see upper left panel in Fig. \ref{fig:sim1}.
It is clearly visible that the image treatment lowers the $snr$ of peaks in the $M_{\rm ap}$-maps of the input catalogue significantly.

We quantify this effect by plotting the number density of $snr$-peaks in the $M_{\rm ap}$-maps, see Fig. \ref{fig:sim2}.
The lower left panel of this figure compares the number density of peaks in the $M_{\rm ap}$-maps of the input catalogue and the output catalogue. 
The image treatment definitely lowers the number density of peaks in the $M_{\rm ap}$-map, the number density of the output catalogue is lower than that of the input catalogue by a factor of two.
We obtain 11 peaks per square degree with a $snr$ larger than four for the input catalogue, but only 6 for the output catalogue (see also Tab. \ref{tab:numberdensity}). 
The ratio between the number density of peaks obtained by analysing the input catalogue and output catalogue can therefore be reduced using more conservative SExtractor parameter settings (like a larger number of contiguous pixels).

Faint galaxies have intrinsically the same size as the PSF or smaller.
Hence, the observed galaxy images are composed of only a few pixels and the ellipticity determination via quadrupole moments of the surface brightness is extremely noisy as is the PSF correction which is also calculated by means of the quadrupole moments.  
The reduction of the $snr$ of peaks and the number density of peaks therefore follows from the fact that shear information is destroyed due to the noisiness of galaxy images and the subsequent extremely noisy correction process.

Very long exposure times with ground-based telescopes would not increase the number density of background galaxies appreciably because additional fainter galaxies are much smaller then the PSF so that one cannot measure the pre-seeing surface brightness properly.
For future large weak lensing surveys it is therefore not efficient to propose for a few very deep images to perform a weak lensing analysis compared to many shallower images.
%
\subsection{The effect of weighting }
\begin{figure}
\centering
\resizebox{\hsize}{!}{\includegraphics[width=\textwidth]{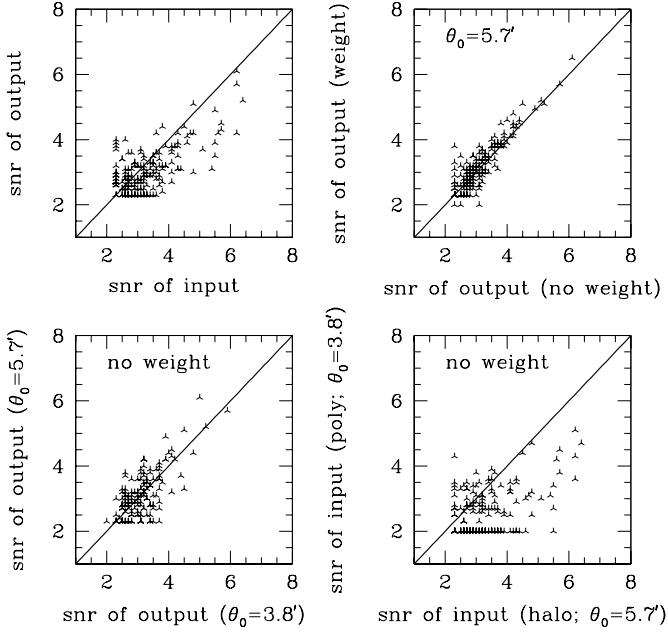}}
\caption{Comparison of the signal-to-noise ratio of peaks in the $M_{\rm ap}$-maps for different analyses.
\textbf{upper left: }Comparison between input catalogues and output catalogues (after image treatment).
\textbf{upper right: }Comparison between weighting and no weighting (for output catalogues).
\textbf{lower left: }Comparison between two different halo-filter radii.
\textbf{lower right: }Comparison between halo-filter and polynomial filter.
\label{fig:sim1}
}
\end{figure}

\begin{figure}
\centering
\resizebox{\hsize}{!}{\includegraphics[width=\textwidth]{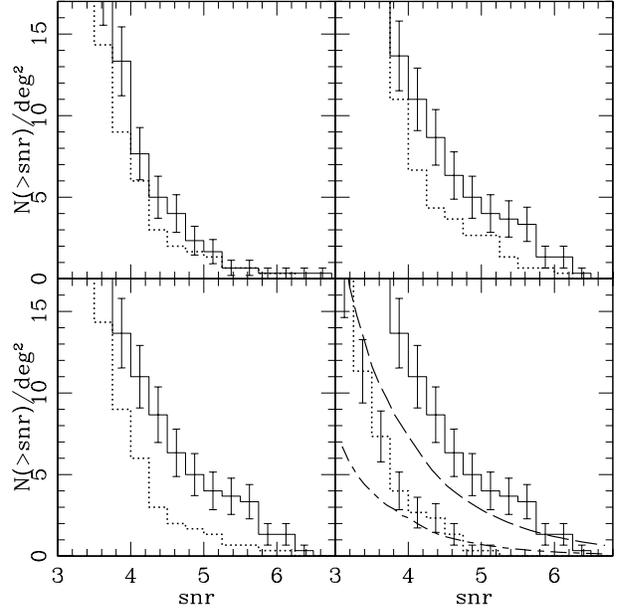}}
\caption{Number density of peaks in the $M_{\rm ap}$-maps per square degree with a signal-to-noise ratio larger than $snr$ obtained from ray-tracing through N-body simulations.
\textbf{Upper left: } dotted histogram: number density of peaks in the $M_{\rm ap}$-maps of output catalogue ($\theta_0=5\myarcmin 7$; no weighting).
Solid histogram: same but with weighting.
\textbf{Upper right: } solid histogram: number density of peaks in the $M_{\rm ap}$-maps of input catalogue ($\theta_0=5\myarcmin 7$; no weighting).
Dotted histogram: same but with a filter radius of $\theta_0=3\myarcmin 8$.
\textbf{Lower left: } solid histogram: number density of peaks in the  $M_{\rm ap}$-maps of input catalogue ($\theta_0=5\myarcmin 7$; no weighting).
Dotted histogram: same but for the output catalogue.
\textbf{Lower right: } solid histogram: number density of peaks in the $M_{\rm ap}$-maps of input catalogue for the halo-filter ($\theta_0=5\myarcmin 7$; no weighting). 
Dotted histogram: same but for the polynomial filter with $\theta_0=3\myarcmin 8$.
Long dashed line: analytical model with same parameters as simulations (halo-filter). Long-short dashed line: same but for the polynomial filter.
The error bars are due to Poisson statistics obtained from 3 square degrees. 
Because of clarity they are not plotted for every line.
}
\label{fig:sim2}
\end{figure}

To study the effect of weighting (the weighting scheme is explained in Sect. \ref{sec:catalogue}) we apply the aperture mass statistics to the output catalogues without weighting, Eq. (\ref{map}), and with weighting, Eq. (\ref{mapweight}).
It turns out that weighting in the form described in Sect. \ref{sec:catalogue} on average raises the $snr$ of peaks in the $M_{\rm ap}$-map only slightly, see upper right diagram in Fig. \ref{fig:sim1}.
The small influence of weighting is mainly due to the fact that we exclude PSF corrected ellipticities of more than 0.8.  
We plot the number density of peaks in the $M_{\rm ap}$-maps with and without weighting in the upper left diagram of Fig. \ref{fig:sim2}.
The number density increases slightly using weighting.
We obtain 6 peaks per square degree with a $snr$ larger than four for the output catalogue without weighting, but 7.5 for the same catalogue with weighting.
%
%
%
%
%
%
%
%
%
\section{Border effects }
\label{sec:bordereffect}
%
A maximum $snr$ for a given filter function $Q$ is attained when the centre of a radially symmetric mass concentration and the circular aperture centre coincide.
However, if the aperture lies partly outside a field, the $snr$ is lowered due to the fact that the number of galaxies from which the shear is estimated decreases. 
To avoid these border effects, the distance $x$ of the centre of the circular aperture from the edge of the field must be larger than the aperture radius.
The resulting subfield would be significantly smaller, even for wide-field images
(the radius of the halo-filter, for which one obtains the maximum number density of NFW-halos is $\theta_0 \approx 6^\prime$, the subfield of a $30^\prime\times 30^\prime$ field would then be $18^\prime\times 18^\prime$, consequently $64\,\%$ smaller).

At this point it is worth mentioning that the condition of a compensated filter function $U$ [see Eq. (\ref{compensate})] is no longer valid if the circular aperture lies partly outside the field, so that the $M_{\rm ap}$-value measured is not fully related to the projected mass distribution, $\kappa$.
Note, however, that the goal of this method is to find mass concentrations and not to get detailed information about the mass distribution.
%
%
%
\begin{figure}
\centering
\resizebox{\hsize}{!}{\includegraphics[width=\textwidth,clip]{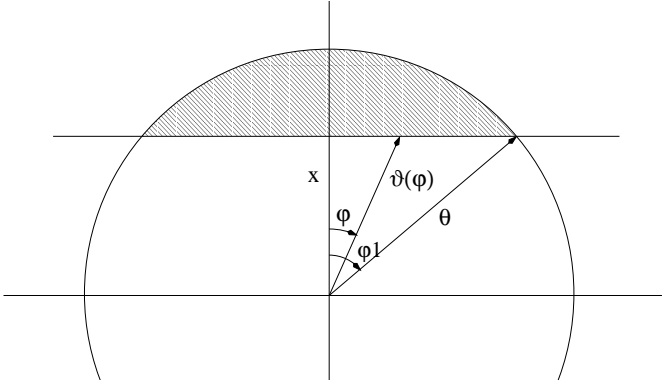}}
\caption{The figure displays a part of the circular aperture with radius $\theta$, which lies partly outside a field (shaded area). The parameter $x$ denotes the distance between the aperture centre and the edge of the field, $\vartheta(\varphi)$ is the parameterised radius and $\varphi_1$ is the integration limit. }
\label{fig:mapmodzeichnung}
\end{figure}
\subsection{Border effects for an SIS profile  }
\begin{figure}
\centering
\resizebox{\hsize}{!}{\includegraphics[width=\textwidth]{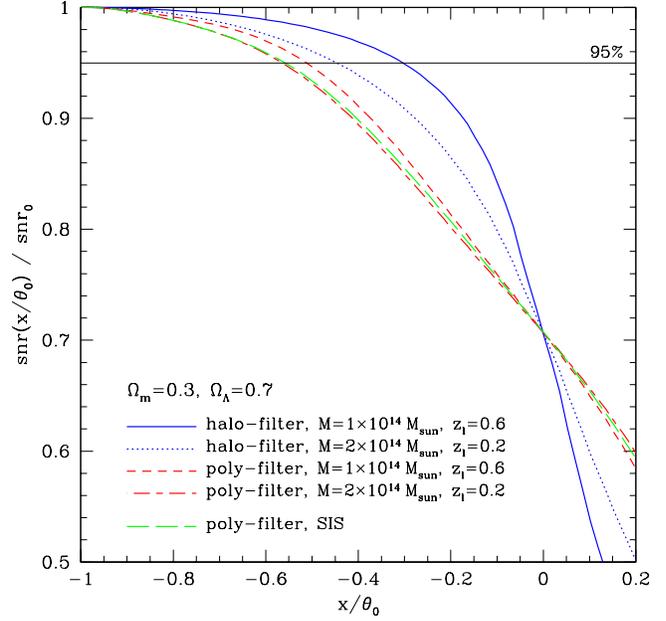}}
\caption{Displayed is the relative change of the $snr$ if a circular aperture lies partly outside a field and a cluster with NFW-profile is positioned at the centre of the aperture for different filter types, halo masses and halo redshifts.
The same is displayed for an SIS-profile, where the polynomial filter is used.
The quantity $x$ is the distance of the aperture centre to the edge and $\theta_0$ is the radius of the aperture. 
Negative $x$: aperture centre inside the field, positive $x$: aperture centre outside the field.}
\label{fig:mapmod}
\end{figure}

\begin{figure}
\centering
\resizebox{\hsize}{!}{\includegraphics[width=\textwidth]{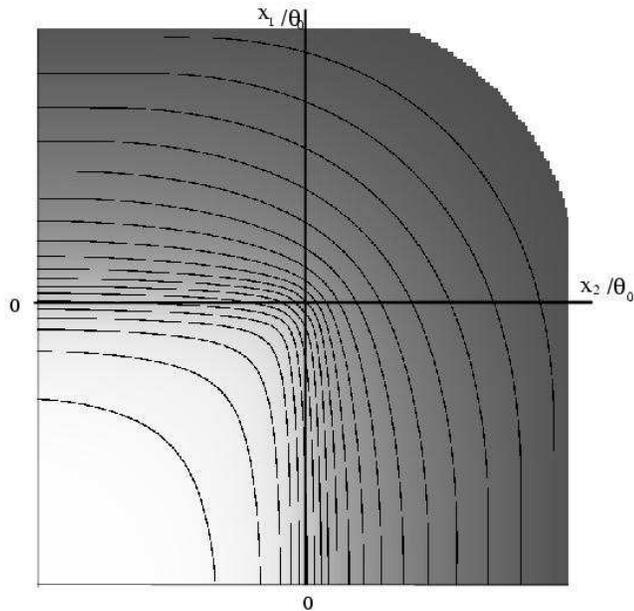}}
\caption{Relative change of the $snr$ if the circular aperture reaches the corner of a field. 
Assumed is an SIS-profile and the halo-filter. 
The contour levels from lower left to upper right are $95\%$ to $5\%$ of the full aperture $snr$.
The centre of the cross indicates the corner of the field, where $x_1/\theta_0=x_2/\theta_0=0$; 
$\theta_0$ is the aperture radius.
}
\label{fig:cornereffectMi}
\end{figure}

As an example we calculate how the $snr$ of a cluster with a profile of a singular isothermal sphere (SIS) centred in the circular aperture changes as a function of the distance $x$ of the aperture centre to the edge of the field (the calculations below are done with the polynomial filter function, because in this case the solution is analytic).
We introduce the parameterised radius $\vartheta(\varphi)=x/\cos\,\varphi$ for $\varphi\in[-\varphi_1,\varphi_1]$ with $\varphi_1=\arccos(x/\theta)$, see Fig. \ref{fig:mapmodzeichnung}.  
With this we then obtain for the signal
\be
M_{\rm ap}=2(I_{M1}+I_{M2}),
\ee
where
\be
I_{M1}=\int_0^{\varphi_1} \dd\varphi \int_0^{\vartheta(\varphi)} \dd\vartheta^\prime\, \vartheta^\prime\,\gamma_{\rm t}(\vartheta^\prime)Q(\vartheta^\prime)
\ee  
and
\be
I_{M2}=\int_{\varphi_1}^\pi \dd\varphi \int_0^\theta \dd\vartheta^\prime\, \vartheta^\prime\,\gamma_{\rm t}(\vartheta^\prime)Q(\vartheta^\prime).
\ee
We solve the integral for an SIS-profile and the filter function (\ref{filter}).
The tangential shear profile of an SIS reads
\be
\gamma_{\rm t}=\frac{\theta_{\rm E}}{2\theta},
\ee
where $\theta_{\rm E}$ is the Einstein radius.
We obtain,
\be
I_{M1}=\frac{x\,\theta_{\rm E}}{40\pi \theta^6}\left\{\theta^2(14\,\theta^2-9x^2)\sqrt{1-\frac{x^2}{\theta^2}}+f_-+f_+\right\},
\ee
with
\be
f_\pm=(\pm20\,\theta^2 x^2\mp9x^4)\ln\left\{\cos\left(\frac{1}{2}\varphi_1\right)\pm\sin\left(\frac{1}{2}\varphi_1\right)\right\}
\ee
and
\be
I_{M2}=\frac{2\,\theta_{\rm E}}{5\pi\theta}(\pi-\varphi_1).
\ee
For the $M_{\rm ap}$ signal of the whole aperture we obtain, $M_{\rm ap}=4\theta_{\rm E}/(5\,\theta)$.
For the noise we find,
\be
\sigma_{\rm Map}^2=2\frac{\sigma_\epsilon^2}{2n}(I_{\sigma1}+I_{\sigma2}),
\ee
where
\be
I_{\sigma1}=\int_0^{\varphi_1}\dd\varphi \int_0^{\vartheta(\varphi)} \dd\vartheta^\prime\, \vartheta^\prime\,Q^2
\ee
and
\be
I_{\sigma2}=\int_{\varphi_1}^\pi \dd\varphi \int_0^\theta \dd\vartheta\,\vartheta\, Q^2.
\ee
The noise is independent of the signal. 
We obtain
\be
I_{\sigma1}=\frac{x}{175\pi^2\,\theta^{11}}\,p\,\sqrt{1-\frac{x^2}{\theta^2}},
\ee
with
\be
p=55\,\theta^8+90\,\theta^6x^2+296\,\theta^4x^4-592\,\theta^2x^6+256\,x^8
\ee
and 
\be
I_{\sigma2}=\frac{3}{5\pi^2\theta^2}(\pi-\varphi_1).
\ee
For the $snr$ and its dependence on the distance $x$ from the edge of the field, we only have to calculate $M_{\rm ap}/\sigma_{\rm Map}$.
The result is shown in Fig. \ref{fig:mapmod}.
As expected, the $snr$ is reduced by a factor of $1/\sqrt{2}$ if the centre of the aperture lies at the edge.
Having only half of the aperture is equivalent to having the half of the galaxies in the whole aperture. 
%
%
We also calculate, for an SIS-profile and the halo-filter, how the $snr$ changes if the circular aperture reaches the corner of a field.
The result is shown in Fig. \ref{fig:cornereffectMi}.
%
%
\subsection{Border effects for an NFW profile }
We calculate numerically how these border effects affect the $snr$ of a halo with NFW-profile positioned in the centre of the aperture.
The results are also shown in Fig. \ref{fig:mapmod}.
We conclude that if one uses the halo-filter the $snr$ drops significantly (more than $5\%$) only if the centre of the aperture is closer to the border of a field than $0.4 \times \theta_0$.
\subsection{Influence of border effects on the number density of peaks in the $M_{\rm ap}$-map}
\begin{figure}
\centering
\resizebox{\hsize}{!}{\includegraphics[width=\textwidth]{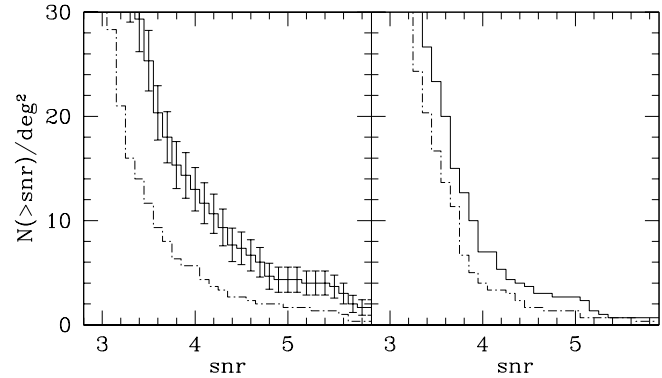}}
\caption{Influence of the border effects on the number density of halos.
Displayed is the comparison of the number density for twelve $30^\prime\times 30^\prime$ images (solid lines) and for 300 $6^\prime\times 6^\prime$ images (dashed lines) of the input catalogue.
Left: halo-filter radius $\theta_0=5\myarcmin 7$.
Right: halo-filter radius $\theta_0=3\myarcmin 8$.
For a better comparison the error bars are only plotted once. 
}
\label{fig:sim3}
\end{figure}

Using the filter function (\ref{Mfilter}), the optimal aperture radius to detect a maximum number of NFW-halos is theoretically $\theta_0=6^\prime$.
As each VLT/FORS1 field only covers $6\myarcmin 8 \times 6\myarcmin 8$, the optimal radius has the same size as the entire VLT field.
We have seen that the $snr$ decreases if the aperture is partly outside a field.
We now determine how this affects the expected number density of halos.
Therefore, we split each of the 12 fields of the numerical simulations into 25 subfields all having a size of $6^\prime\times 6^\prime$, obtaining 300 VLT-field-sized images.
We apply the $M_{\rm ap}$-statistics to these subfields for two filter radii ($\theta_0=3\myarcmin 8$ and $\theta_0=5\myarcmin 7$) and find that the number density of peaks drops by a factor of $\sim$ two compared to the large fields, see Fig. \ref{fig:sim3} and Table \ref{tab:numberdensity}. 
%
%
%
\subsection{Influence of border effects on the number density of noise peaks}
\begin{figure}
\centering
\resizebox{\hsize}{!}{\includegraphics[width=\textwidth]{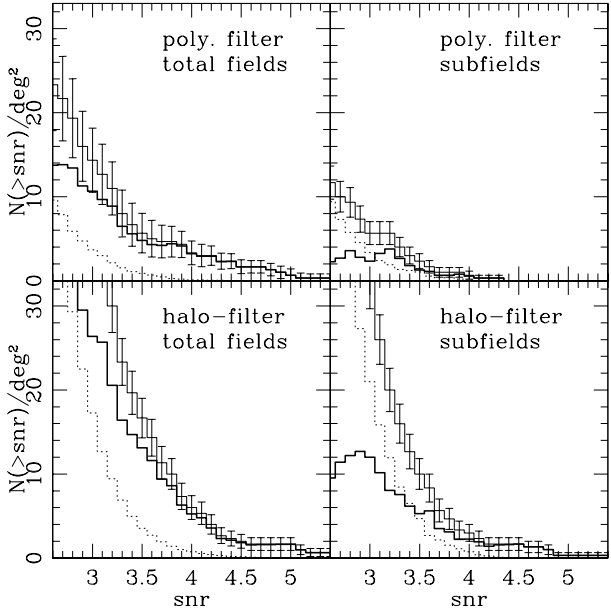}}
\caption{Influence of the border effects on the number density of all peaks and noise peaks.
Displayed is the comparison of the number density for twelve $30^\prime\times 30^\prime$ images (total fields) and for 300 $6^\prime\times 6^\prime$ images (subfields) of the output catalogue for two different filters (filter radius in all cases $\theta_0=5\myarcmin 7$).
Solid lines with error bars: total number density of peaks in the $M_{\rm ap}$-maps. 
Dotted lines: number density of noise peaks, resulting from randomisation of the orientation of background galaxies.
Bold solid lines: difference between total number density of peaks and noise peaks.     
}
\label{fig:real_split} 
\end{figure}

In this section we study the influence of the border effects on the number density of noise peaks and the resulting number density of real peaks in weak lensing maps of small size.
Noise peaks and real peaks are defined as in section \ref{sec:noisepeaks}.
In Fig. \ref{fig:real_split} we compare the number density of all peaks, noise peaks and the resulting real peaks, for twelve $30^\prime\times 30^\prime$ images (total fields) and for 300 $6^\prime\times 6^\prime$ images (subfields) of the output catalogue for two different filter (filter radius in all cases $\theta_0=5\myarcmin 7$).
In Tab. \ref{tab:numbernoise} two examples are shown.

The tremendous influence of the border effects on the number density of real peaks (difference between total number and number of noise peaks) can be seen in Fig. \ref{fig:real_split}.
The total number of peaks in the subfields drops and the number of noise peaks rises compared to the total fields.
The number of real peaks is therefore significantly smaller.
%
%
\section{Weak lensing analysis of the VLT data} 
\label{sec:VLT}
\subsection{The data}
\label{sec:data}
For the current work we observed 50 uncorrelated fields with the FORS1 camera at the VLT (UT1, ANTU) at Paranal.
These fields have been used before for a cosmic shear analysis \citep{mvm01}.
The observations were taken under optimal conditions (average seeing $\sim 0 \myarcsec 63$) in $I$-band during the Period 63 (March 1999 to September 1999). 
The FORS1 camera is equipped with a $2048 \times 2048\, {\rm pixel}$ CCD chip with a pixel size of $0.24\,\mu{\rm m}$
which corresponds to $\sim 0 \myarcsec 2$ in the standard mode of the instrument.
Consequently, each field covers $6\myarcmin 8 \times 6\myarcmin 8$. 
Altogether the 50 fields cover $\sim 0.64\, {\rm deg}^2$.
The total observing time for each field was 36 minutes, which resulted in a limiting magnitude of $I_{\rm AB}\approx24.5$, corresponding to a mean galaxy number density of $26\,{\rm arcmin}^{-2}$ in the final weak lensing catalogue.
The expected mean redshift of the lensed sources is $\langle z \rangle \approx 1$ .
The fields represent 50 independent lines-of-sight chosen in such a way that they are neither biased towards overdense or underdense regions, nor bright stars or other very luminous objects are within or close to the field, in order to have enough background galaxies for the weak lensing analysis.
For the following analysis the VLT data was already reduced.
For further information about the observations, selection criteria of the fields and data reduction we refer to \cite{mvm01}.
\subsection{Catalogue creation and ellipticity correction}
\label{sec:catalogue}
\begin{figure}
\centering
\resizebox{\hsize}{!}{\includegraphics[width=\textwidth]{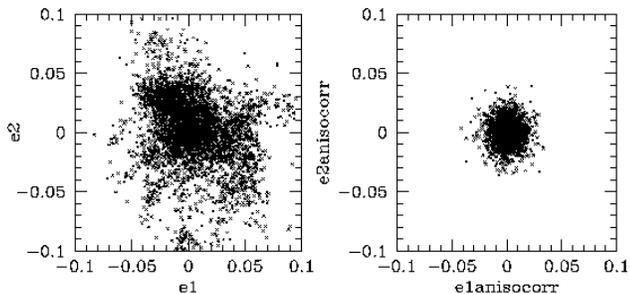}}
\caption{Stellar ellipticities for all VLT fields before (left) and after (right) PSF anisotropy correction.}
\label{fig:psf}
\end{figure}

In this section we briefly elucidate our catalogue creation, catalogue filtering, PSF correction and weighting scheme.  
\\
\textbf{Raw catalogue. }
SExtractor is used to create two primary catalogues of all objects in the $I$-band image which consist of at least $N=3$ and $N=5$ contiguous pixels (SExtractor parameter `DETECT\_MINAREA') with a flux greater than the $k=1\sigma$ and $k=2\sigma$ sky level noise (SExtractor parameter `DETECT\_THRESH') for all 50 VLT fields.
\\  
\textbf{First catalogue filtering.  }
All objects for which problems concerning the determination of shape or position occur are rejected
(e.g., objects near the border, with negative total flux, with negative $Q_{11}+Q_{22}$ (see below), or with negative semi major and/or semi major axis).
Two catalogues of 107100 objects for $N=3,\,k=1$ and 79400 objects for $N=5,\,k=2$ remain.
\\  
\textbf{Second catalogue filtering.  }
The raw background galaxy catalogues are selected from objects with SExtractor isophotal magnitude $I > 18.5$ and a half-light radius which is larger than that measured for stars.
The resulting catalogues contain altogether $\sim 72100$ objects with $N=3,k=1$ and $\sim 52900$ objects with $N=5,k=2$.
\\  
\textbf{The principle of PSF correction.  }
The shape of galaxies is influenced by the anisotropic PSF. 
In order to obtain a correct estimate of the shear $\gamma$ from the observed ellipticity of galaxies $e^{\rm obs}$, \cite{ksb95} developed the so-called KSB algorithm.
The algorithm relates the observed ellipticities $e^{\rm obs}$ to the sheared source-ellipticities and provides an unbiased estimator of the shear.
The correction is calculated on the second brightness moments $Q_{ij}$ of a galaxy with surface brightness $I(\vek{\theta})$.
The quantity $Q_{ij}$ is defined by
\begin{equation}
Q_{ij}=\int {\rm d}^2\theta \,(\theta_i-\bar\theta_i)(\theta_j-\bar\theta_j)\,I(\vek{\theta})\,W\left(\left|\vek{\theta}-\vek{\bar\theta}\right|^2\right),
\label{quadro}
\end{equation}
where $W$ is a window function with a smoothing scale $r_{\rm g}$ (connected to the object size), which suppresses the photon noise of the objects profile at large radii, and $\bar\theta$ is the centre of the surface brightness.
The ellipticity is defined as
\begin{equation} 
e:=\frac{Q_{11}-Q_{22}+2{\rm i}Q_{12}}{Q_{11}+Q_{22}}.
\label{elli}
\end{equation}
Assuming that the intrinsic orientation of galaxies is random, the relation between $\gamma$ and $e^{\rm obs}$ in the weak lensing regime reads
\be
\gamma= (P^{\rm g})^{-1}(e^{\rm obs}-P^{\rm sm}q^*),
\ee
where $P^{\rm g}$ is the pre-seeing shear polarisability which depends on the smear and shear polarisability tensors $P^{\rm sm}$ and $P^{\rm sh}$, and the stellar smear and shear polarisability tensors $P^{{\rm sm}*}$ and $P^{{\rm sh}*}$. The tensor $P^{\rm g}$ distorts the galaxy ellipticity to its true value.
Both quantities are calculated by means of the observable $Q_{ij}$.
The quantity $q^*$ is the stellar ellipticity (due to the PSF-anisotropy) and is calculated from the raw stellar ellipticity $e^*$, $q^*=(P^{\rm sm *})^{-1}e^*$.
\\
\textbf{PSF-anisotropy correction.  }
The stars which are used for the PSF-anisotropy correction are selected by plotting magnitude against half-light radius ($mag-r_{\rm h}$ plot).
All stars have the same half-light radius and therefore show up as a vertical branch in this plot.
Stars which have a magnitude of $mag=0.5$ lower than the saturated stars and which are well above the crowded faint magnitude area which contains a mixture of faint stars, galaxies and noise detections are selected.
Using this sample of stars, a third-order two-dimensional polynomial fit with $3.5\sigma$-clipping of the stellar ellipticities $e^*$ (intrinsically round objects are distorted due to the anisotropic part of the PSF and are thus elliptical) as a function of position is performed. 
With this, the quantity $q^*=(P^{\rm sm *})^{-1}e^*$ at the position of the galaxies is calculated.
Fig. \ref{fig:psf} displays the stellar ellipticities for all VLT fields before and after the PSF anisotropy correction.
%
%
\\
\textbf{Calculation of $P^{\rm g}$.  }
The diagonal elements of the $P^{\rm g}$ tensor are dominant by a factor of 10 compared to the off-diagonal elements and they are approximately equal, so that we can estimate $P^{\rm g}$ by $P^{\rm g}_{\rm s}\mathbbm{1}$, with $P^{\rm g}_{\rm s}=0.5\,{\rm trace}[P^{\rm g}]$ \citep{ewb01}.
The stellar smear and shear polarisability tensors $P^{{\rm sm}*}$ and $P^{{\rm sh}*}$ are calculated for different smoothing scales $r_{\rm g}$.
As $P^{\rm g}$ depends on these quantities we calculate $P^{\rm g}$ according to the galaxy size.
\\
\textbf{Third catalogue filtering.  }
All objects having an ellipticity (after PSF correction) of more than 0.8 are rejected.
The final catalogues consist of 56800 galaxies for $N=3,\,k=1$ and 42200 galaxies for $N=5,\,k=2$, resulting in an average number density of $n=26.3/{\rm arcmin}^2$ and $n=19.5/{\rm arcmin}^2$, respectively.
The galaxy ellipticity dispersion of the final catalogues is on average $\sigma_e=0.4$. 
In Table \ref{SumTab} seeing condition and galaxy number density of the final catalogues are listed for all 50 VLT fields.
\\
\textbf{Weighting.  }
Since the corrected galaxy ellipticities are very noisy, a weighting scheme according to the noise level is introduced. 
For each galaxy the next twelve neighbours are identified in the $mag-r_{\rm h}$ plane and the variance $\sigma_e^2$ of the ellipticity distribution of the sub-sample is calculated, see \citet{ewb01}.
The variance $\sigma_e^2$ gives an indication of the noise level of these galaxies.
According to $\sigma_e^2$ we then determine the weighting factor $w$ as,  
\be
w=1/\sigma_e^2.
\label{gewichtung}
\ee
\subsection{Number density and seeing}
The signal-to-noise ratio of a halo detection using the aperture mass statistics is proportional to the square root of the number density of background galaxies, and the expected number density of $M_{\rm ap}$-peaks strongly depends upon the $snr$.
Seeing conditions are therefore crucial.
For this work we have a large data set of VLT fields, all taken under different seeing conditions, so we briefly present the dependence of galaxy number density on seeing and SExtractor parameter settings. 
 
\begin{figure}
\centering
\resizebox{\hsize}{!}{\includegraphics[width=\textwidth]{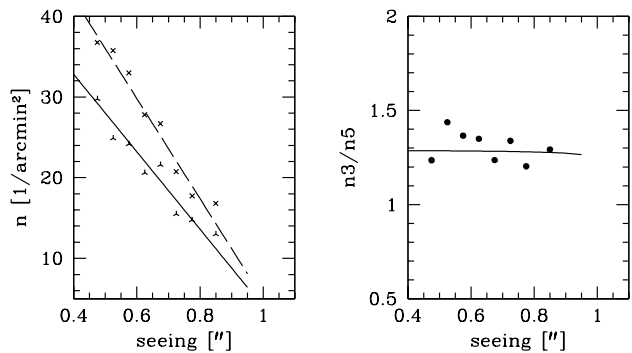}}
\caption{The diagram on the left displays the number density of galaxies per square arcminutes and its dependence on seeing in arcseconds for two different SExtractor parameter settings. 
Solid line: fit to the binned data for the parameter settings: $N=5, k=2$; dashed line: $N=3,k=1$. 
The diagram on the right displays the ratio between the two number densities caused by the two different SExtractor parameter settings. 
The solid line is the ratio between the two fits.
The number density of galaxies are from the final catalogue.
Data from Table \ref{SumTab}.}
\label{fig:seeing}
\end{figure}

Fig. \ref{fig:seeing} illustrates the dependence of the number density of galaxies used for the weak lensing analysis (final lensing catalogue) on seeing and SExtractor parameter settings (data from Table \ref{SumTab}).
The ratio of the number density between the SExtractor parameter settings $N=3,\,k=1$ and $N=5,\,k=2$ is approximately constant ($n3/n5\approx 1.3$) as a function of seeing.
Using the parameter settings $N=3,\,k=1$ we expect the $snr$ of a cluster detection to be $1.14$ times larger compared to the parameter settings $N=5,\,k=2$. 
However, this is only true if the additional sources are not dominated by noise detections which do not contain shear information.

The ratio of the number density of galaxies between a seeing condition of $0\myarcsec 6$ and $0\myarcsec 9$ is approximately 2.7, which corresponds to a ratio of the $snr$-values of cluster detections of 1.6.
Hence the seeing conditions have a tremendous influence on the expected number density of detectable clusters.
 
We have seen that the galaxy number density strongly depends on the seeing conditions and the SExtractor parameter settings, and the information about the number density of galaxies only makes sense if simultaneously information about source extraction and seeing are given.
%
%
%
%
\subsection{Comparison to numerical simulations}
\label{sec:comparisonvltsim}
In this section the results of the simulations are compared with real data.
Therefore, we apply the $M_{\rm ap}$-statistics to the VLT-images as described in Sect. \ref{sec:mapondata}.
A regular grid with a mesh size of $10^{\prime\prime}\times 10^{\prime\prime}$ is placed over the data fields and the $snr$ is calculated for every grid point.
This is done for the SExtractor parameter settings 3 contiguous pixels $1\sigma$ above the sky background and two different halo-filter radii, $\theta_0=3\myarcmin 8$ and $\theta_0=5\myarcmin 7$.
The number density of background galaxies in the numerical simulations is $n=19\,{\rm arcmin}^{-2}$ and the ellipticity distribution is $\sigma_e=0.34$ compared to $n<26\,{\rm arcmin}^{-2}$ (depending on the redshift of the possible galaxy cluster) and $\sigma_e=0.4$ for the VLT data.
The $snr$ of a halo is proportional to $\sqrt{n}/\sigma_e$. 
As $\sqrt{19}/0.34=12.82\approx 12.74=\sqrt{26}/0.4$ it is legitimate to compare the expected number densities of the numerical simulations with those of the VLT data.

\begin{figure}
\centering
\resizebox{\hsize}{!}{\includegraphics[width=\textwidth]{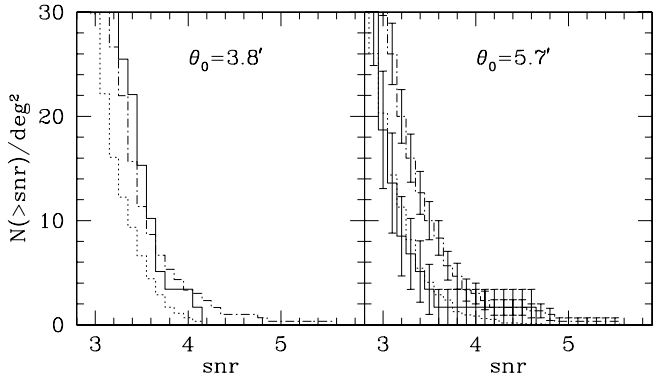}}
\caption{Comparison of number density between simulations, including border effects and image treatment (dot-dashed lines), and VLT data (solid lines).
Dotted lines: noise peaks due to randomisation of background galaxies of all VLT fields. 
}
\label{fig:sim-VLT}
\end{figure}

In Fig. \ref{fig:sim-VLT} we add up all peaks in the $snr$-maps of all VLT fields for the two filter radii, normalise it to one square degree and compare it with the number density of peaks obtained in the same way from the 300 subfields created from the numerical simulations (Sect. \ref{sec:sim}). 
The average number of noise peaks per field above a given threshold is determined by randomising the orientations of the background galaxies 20 times for each field (Fig. \ref{fig:sim-VLT}).
For the small filter radius ($\theta_0=3\myarcmin 8$) the number density of peaks in the VLT fields is (within the errors) in good agreement with the simulations and is significantly above the number density of noise peaks.
For the large filter radius ($\theta_0=5\myarcmin 7$) this is not the case.
The number density of peaks in the simulations is significantly higher compared to the VLT fields and is comparable with that of the noise peaks.
The reason for this could be that more massive clusters (which are better matched by a large filter radius) are present in the simulations (as indicated in Sect. \ref{sec:comp}) than in the VLT fields.
%
%

%
%
%
\section{Cluster candidates}
\label{sec:candidates}
In this section the VLT fields are analysed in detail.
We provide some criteria with which it may be possible to separate peaks in the $M_{\rm ap}$-map resulting from real halos from those peaks resulting from a chance alignment of background galaxies. 
The $M_{\rm ap}$-statistics are applied to the VLT-images using a grid with a mesh size of $3^{\prime\prime}\times 3^{\prime\prime}$.
We use both the polynomial filter (\ref{filter}) and the halo-filter (\ref{Mfilter}) with various filter radii ($\theta_0 \in [1^\prime,6^\prime]$) and two SExtractor parameter settings.
Thereby 30 weak lensing maps ($M_{\rm ap}$-maps) are obtained for each of the 50 VLT fields.
The weak lensing maps are compared with light distribution maps.

The light distribution is calculated on a regular grid, where each grid point contains the total flux within a weighted circular aperture.
The mean light and the standard deviation $\sigma$ are calculated from all weighted total flux values of the grid points.
We optimised the effectiveness of the aperture mass filter function, halo-filter $U$, to detected cluster-sized dark matter halos. 
Assuming that light follows the dark matter distribution, we opt for the weighting function of the halo-filter $U$; obtained via Eq. (\ref{relation}) and Eq. (\ref{Mfilter}).
To calculate the light distribution we use galaxies in the magnitude interval $I\in [16;22]$ and
choose the same aperture radius as for the weak lensing analysis.

If we detect peaks in the weak lensing maps with a signal-to-noise ratio larger than three which coincides with a light overdensity of $2\sigma$ above the mean light or more, we consider these peaks as cluster candidates and perform a more detailed analysis of these candidates.
We also perform a further analysis if the $snr$ of the weak lensing analysis is larger than four. 

In the 50 VLT fields 12 cluster candidates fulfil these conditions (in Fig. \ref{fig:cluster_candidateI} and Fig. \ref{fig:cluster_candidateII} all 12 images of the candidates with $M_{\rm ap}$- and light distribution contours are presented).
In the following, some criteria to substantiate or weaken the presumption of the 12 cluster candidates to be real clusters are presented.
We elaborate only on the analysis done with the optimised halo-filter, because for the relatively small VLT fields, the polynomial filter function is strongly influenced by the border effects (the polynomial filter function places weight at large radii; the maximum is at $0.7\times \theta_0$, see Fig. \ref{fig:UundQ}).
For each field the number of peaks with $snr>3$ for a filter radius of $\theta_0=3\myarcmin 8$ is shown in Table \ref{SumTab}.
\subsection{$snr$-radius plot }
The value of the $snr$-maximum of the cluster candidates is plotted as a function of filter radius, $\theta_0$, for different filter functions, see Fig. \ref{fig:snrKurven}.
For this we neglect the fact that the position of the maxima can vary by a few pixels.
A criterion for a promising cluster candidate is that the $snr$ should be larger than 3, independent of the halo-filter radius in use.
Especially for larger radii (in the case of the small fields at hand a large radius means $\theta_0\approx 3^\prime-5^\prime$) the $snr$ should at least be larger than 3.
Due to this criterion we reject `vlt36' and `vlt60' as cluster candidates.
\subsection{Different SExtractor parameter settings  }
The 50 VLT fields are analysed for two different SExtractor parameter settings, $n=3$ contiguous pixel $k=1\sigma$ above the sky background and a more conservative one, $n=5$, $k=2$. 
With the first parameter settings on the one hand many noisy sources are extracted (many of them excluded by the condition that the half-light radius of sources must be larger than that of stars), which could lower the lensing signal.
With the conservative settings on the other hand, background sources which could contribute to the lensing signal are missed. 
One criterion for a promising cluster candidate is that the $snr$ of the candidate should be larger than $3$ over a large range in radius, independent of the SExtractor parameter settings.
In the case of candidate `vlt77' (see Fig. \ref{fig:snrKurven}) this criterion is not fulfilled.
The $snr$-curve obtained by using the conservative settings is always lower than $2.6$ and is $1\sigma$ or more below the $snr$-curve determined with the other settings (independent of the filter radius).
\subsection{Exclusion of high ellipticities  }
A high $snr$-peak in the $M_{\rm ap}$-map can be caused by a chance alignment of only a small number of galaxies with a high tangential ellipticity, especially if the number density of background sources is low.
Such a statistical fluke can be exposed by restricting the absolute value of the PSF-corrected ellipticity to $|\epsilon|<0.5$.
We analyse the cluster candidates once again using the ellipticity restriction for the SExtractor parameter settings $n=3$, $k=1$ (Fig. \ref{fig:snrKurven}).
The most remarkable example for such a fluke is candidate `vlt79'.
The $snr$ is larger than 4 over a large range of radius, independent of the SExtractor parameter settings.
If the galaxy ellipticity is restricted to $|\epsilon|<0.5$ the signal drops below $2\sigma$ and is therefore rejected.
A further example is `vlt42' which is rejected, too. 
\subsection{Tangential ellipticity-radius plot  }
\begin{figure*}
\centering
\resizebox{\hsize}{!}{\includegraphics[width=\textwidth]{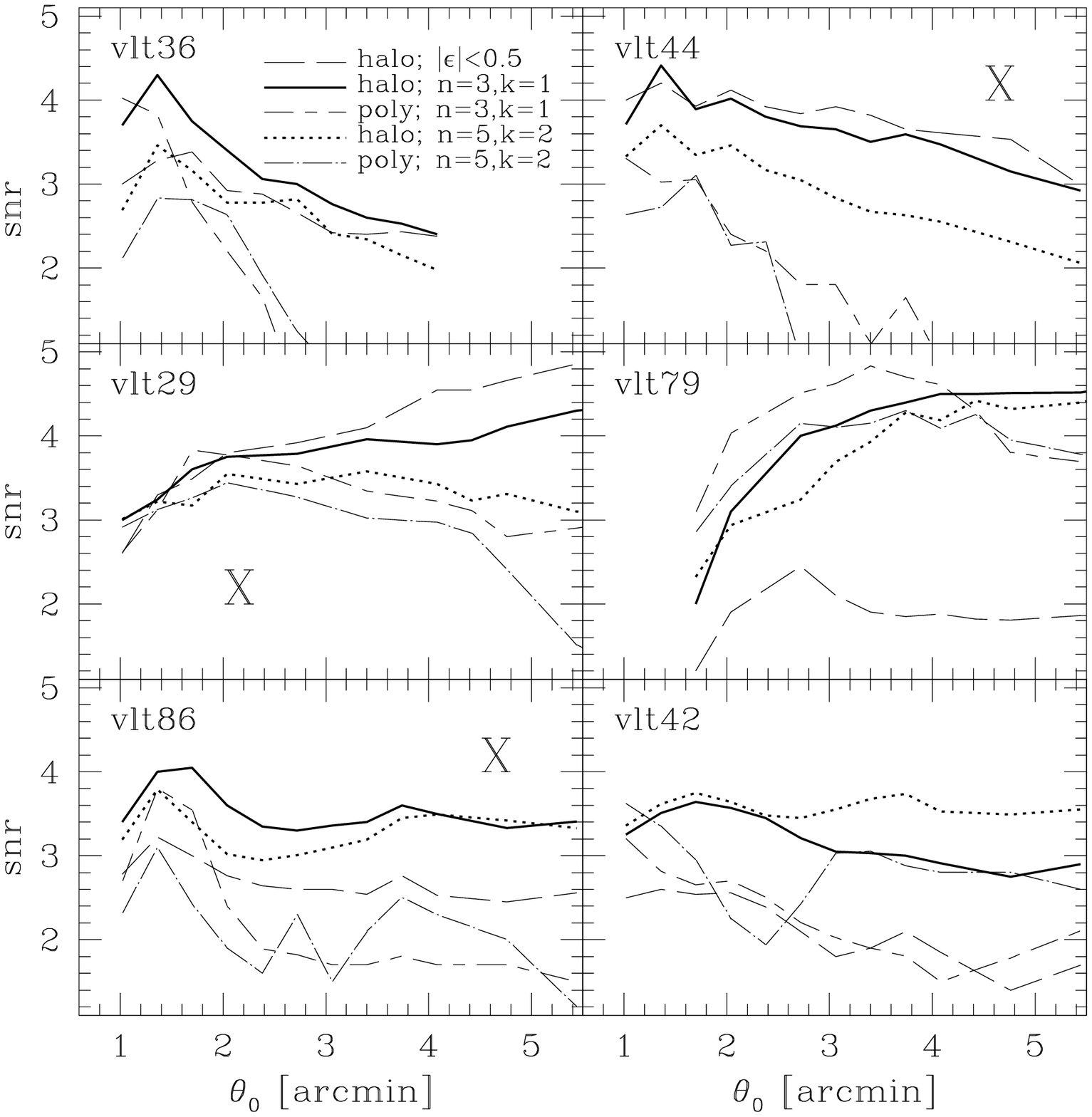}
\includegraphics[width=\textwidth]{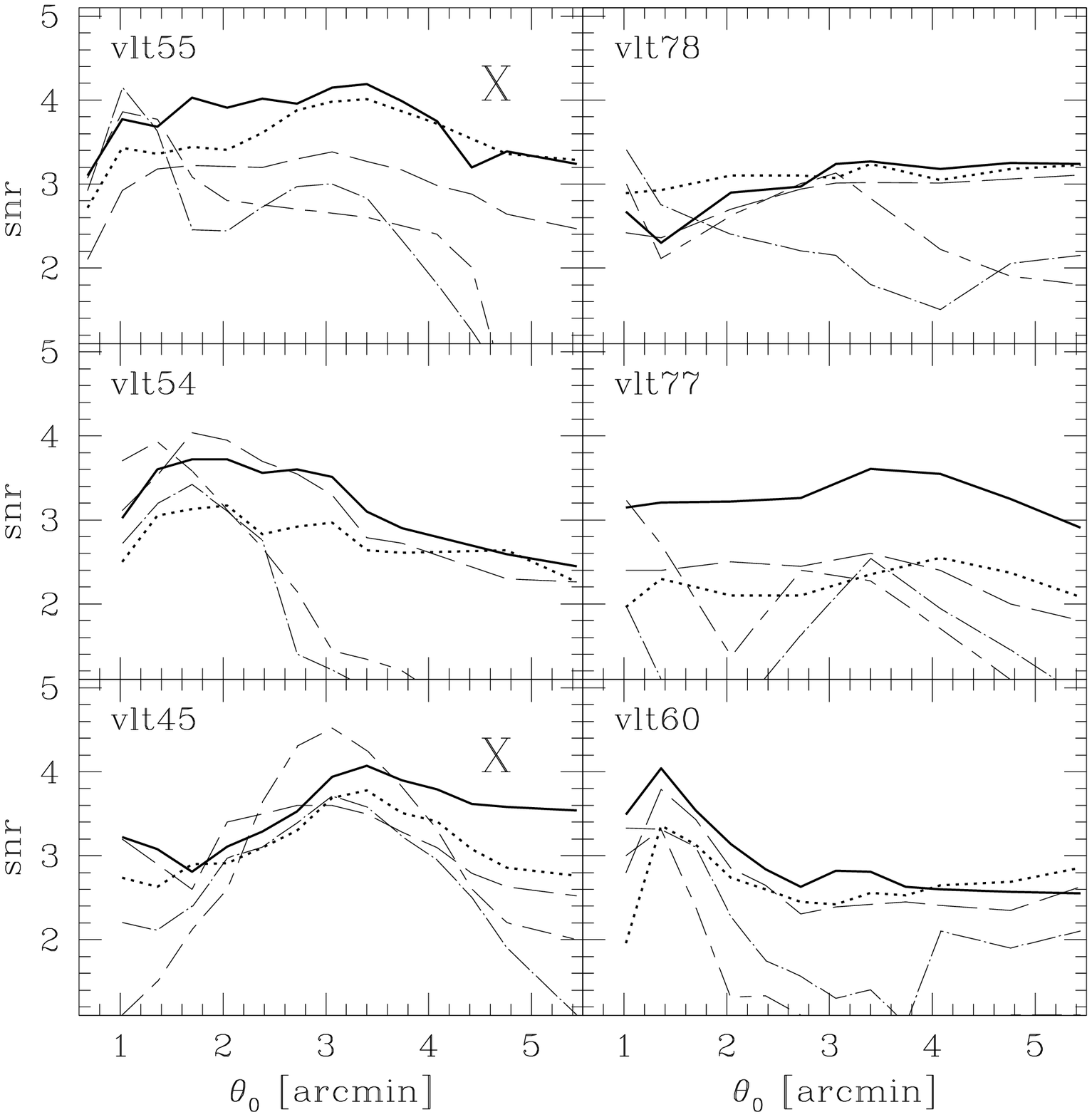}
}
\caption{The detection significance of the shear-selected cluster candidates.
Shown is the maximum signal-to-noise ratio of the cluster candidates in the $M_{\rm ap}$-map and its dependence on filter radius $\theta_0$ for different filter (halo- and polynomial filter function) and SExtractor parameter settings ($n=3$ and $n=5$ contiguous pixel, $k=1\sigma$ and $k=2\sigma$ above the sky background). 
To exclude that the $snr$ of a cluster candidate is dominated only by a small fraction of galaxies with a high ellipticity we also show a $M_{\rm ap}$-analysis for which we exclude ellipticities with $|\epsilon|>0.5$.
The capital X denotes the promising cluster candidates.}
\label{fig:snrKurven}

\centering
\resizebox{\hsize}{!}{\includegraphics[width=\textwidth]{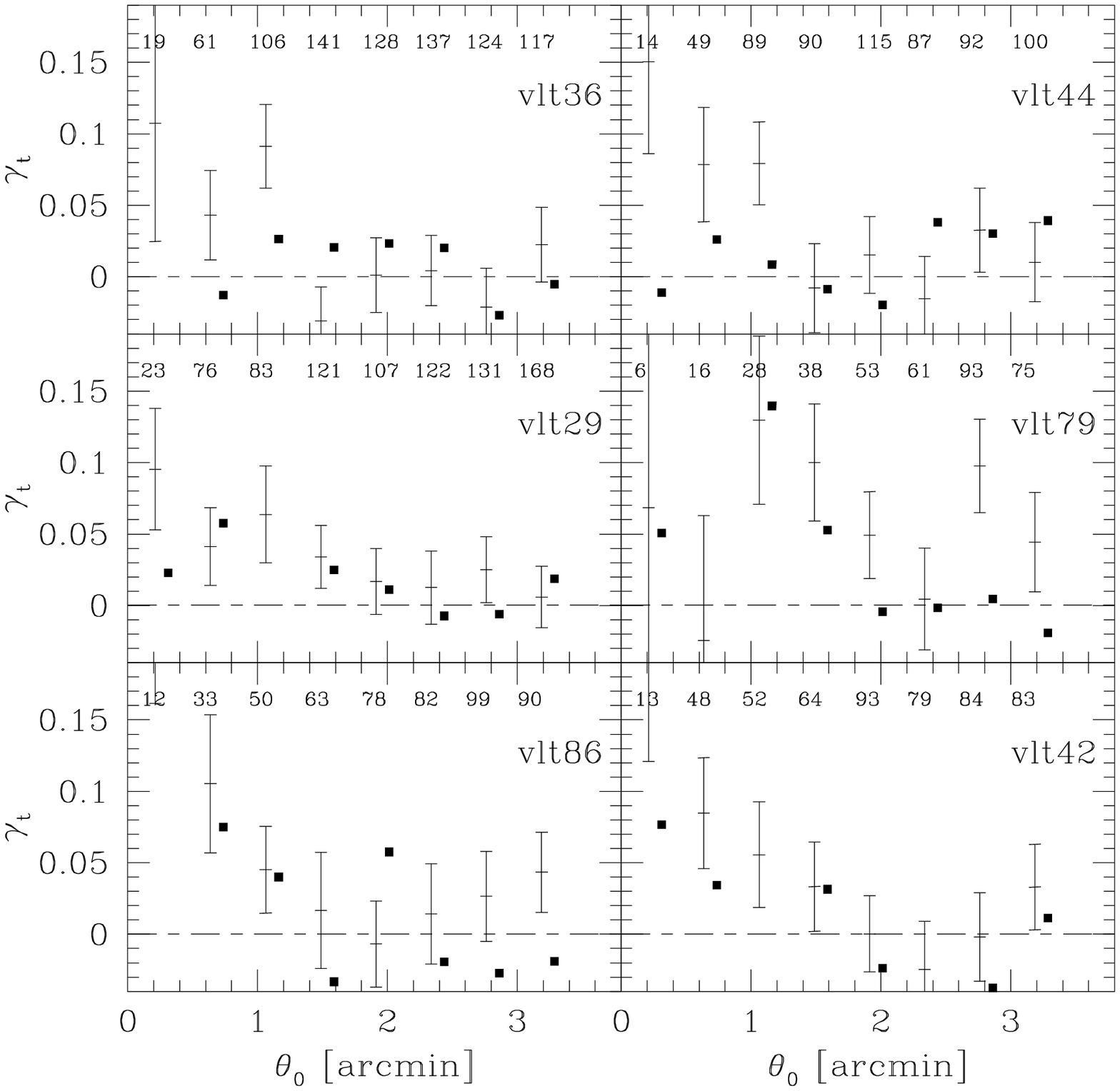}
\includegraphics[width=\textwidth]{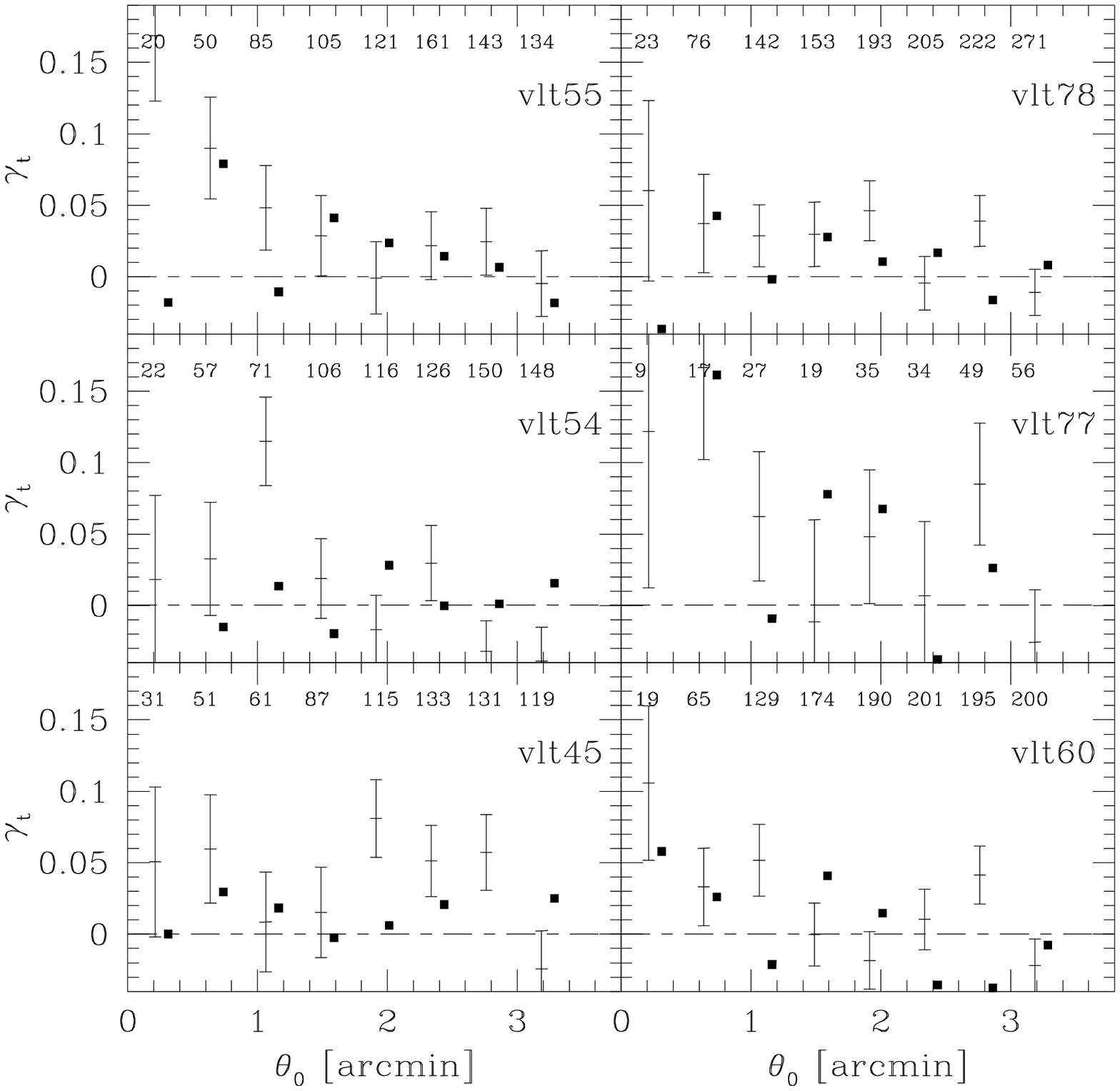}}
\caption{The average tangential shear calculated in rings and its dependence on the distance, $\theta$ (in arcmin), from the $snr$-maximum of the shear-selected cluster candidates.
The error bars are calculated from the cross component of the shear. 
The solid squares without error bars are the cross components of the shear and are shown as a comparison.
The numbers in the upper part of each figure denote the number of galaxies per ring.
Note that some of the cluster candidates are close to the edge of a field.
}
\label{fig:etangential}
\end{figure*}

Another test of the cluster candidates is the tangential ellipticity-radius plot.
The average tangential shear around the $snr$-maximum of a shear-selected cluster candidate is calculated in rings and is plotted against distance from the $snr$-maximum, see Fig. \ref{fig:etangential}.
Assuming a relaxed cluster (an SIS for example), the tangential shear profile should follow roughly a $1/\theta$-relation.
If there are only one or two rings which cause the shear signal in the $M_{\rm ap}$-map, then a cluster candidate is rejected.
An example is the candidate `vlt54'.
\subsection{Five promising cluster candidates }
\begin{figure*}
\centering
\includegraphics[width=16cm]{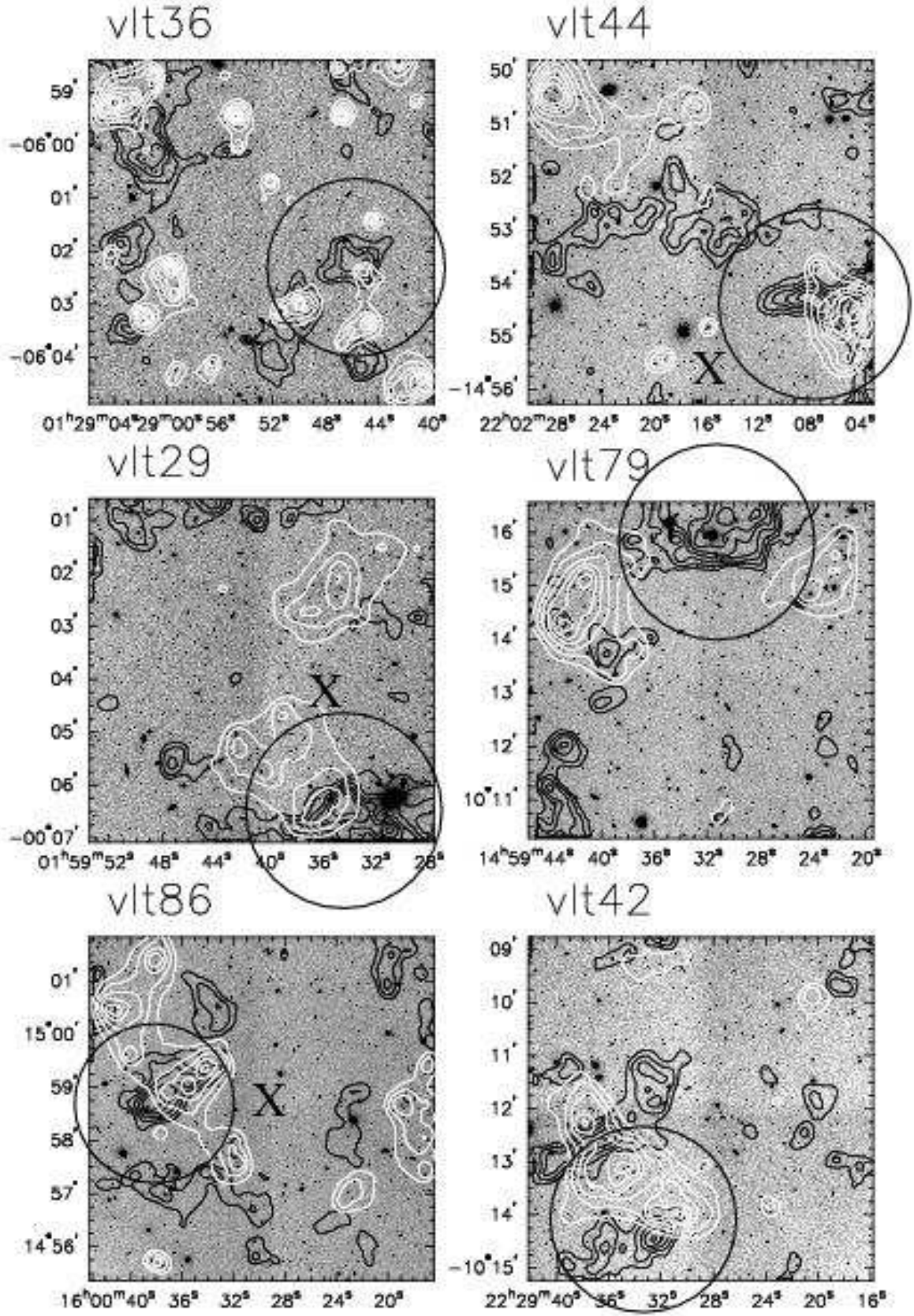}
\caption{VLT fields of all cluster candidates (indicated by circles). 
The $M_{\rm ap}$-contours are black ($snr$-values are 1.5, 2.0, 2.5, 3.0, 3.5, 4.0) and the light distribution contours are white ($snr$-values are 1, 1.5, 2, 2.5, 3, 3.5, 4).
The capital X denotes fields with promising cluster candidates. }
\label{fig:cluster_candidateI}
\end{figure*}

\begin{figure*}
\centering
\includegraphics[width=16cm]{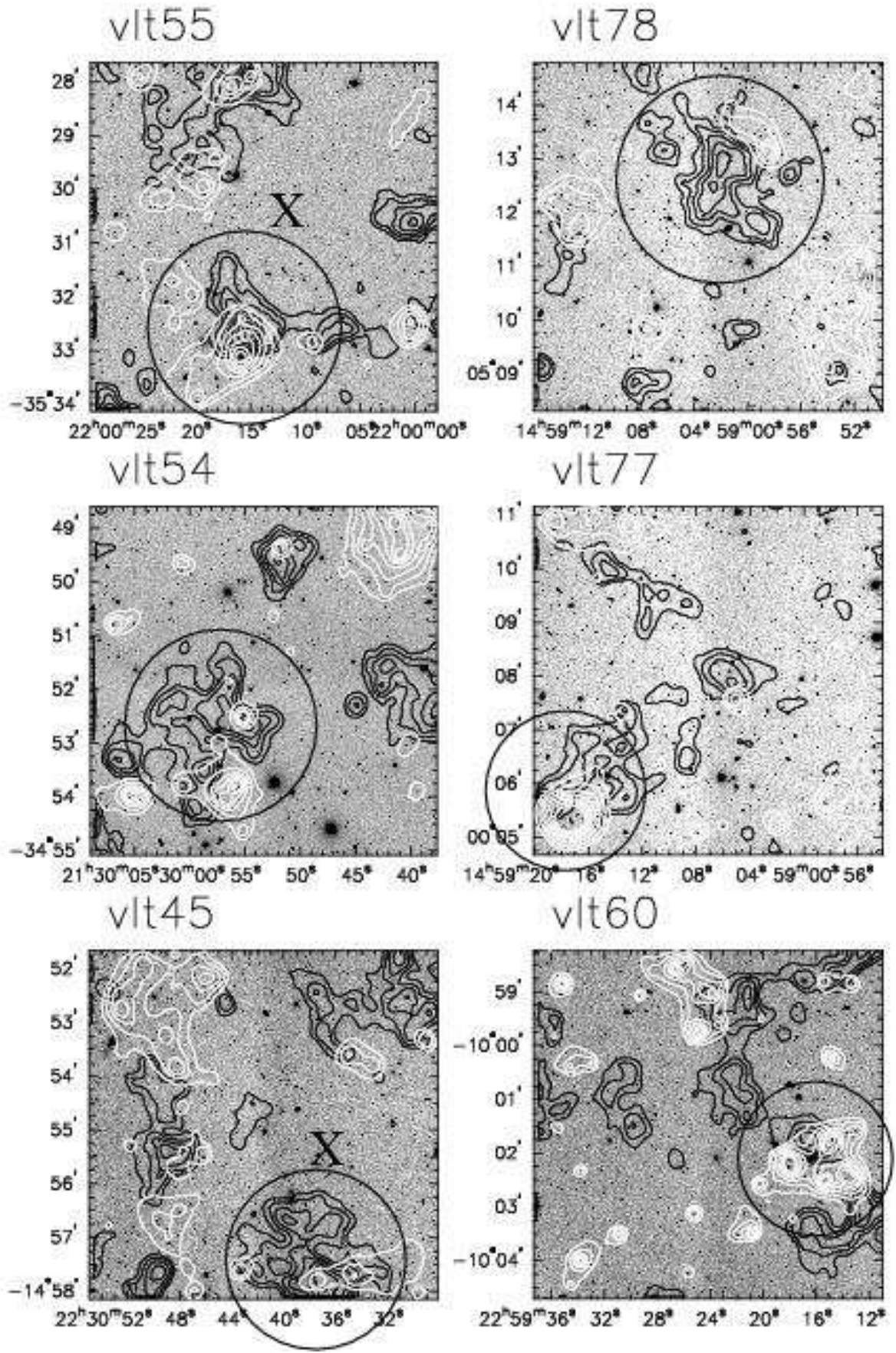}
\caption{As Fig. \ref{fig:cluster_candidateI}, for the other six fields with cluster candidates. }
\label{fig:cluster_candidateII}
\end{figure*}

\begin{figure*}
\centering
\includegraphics[width=16cm]{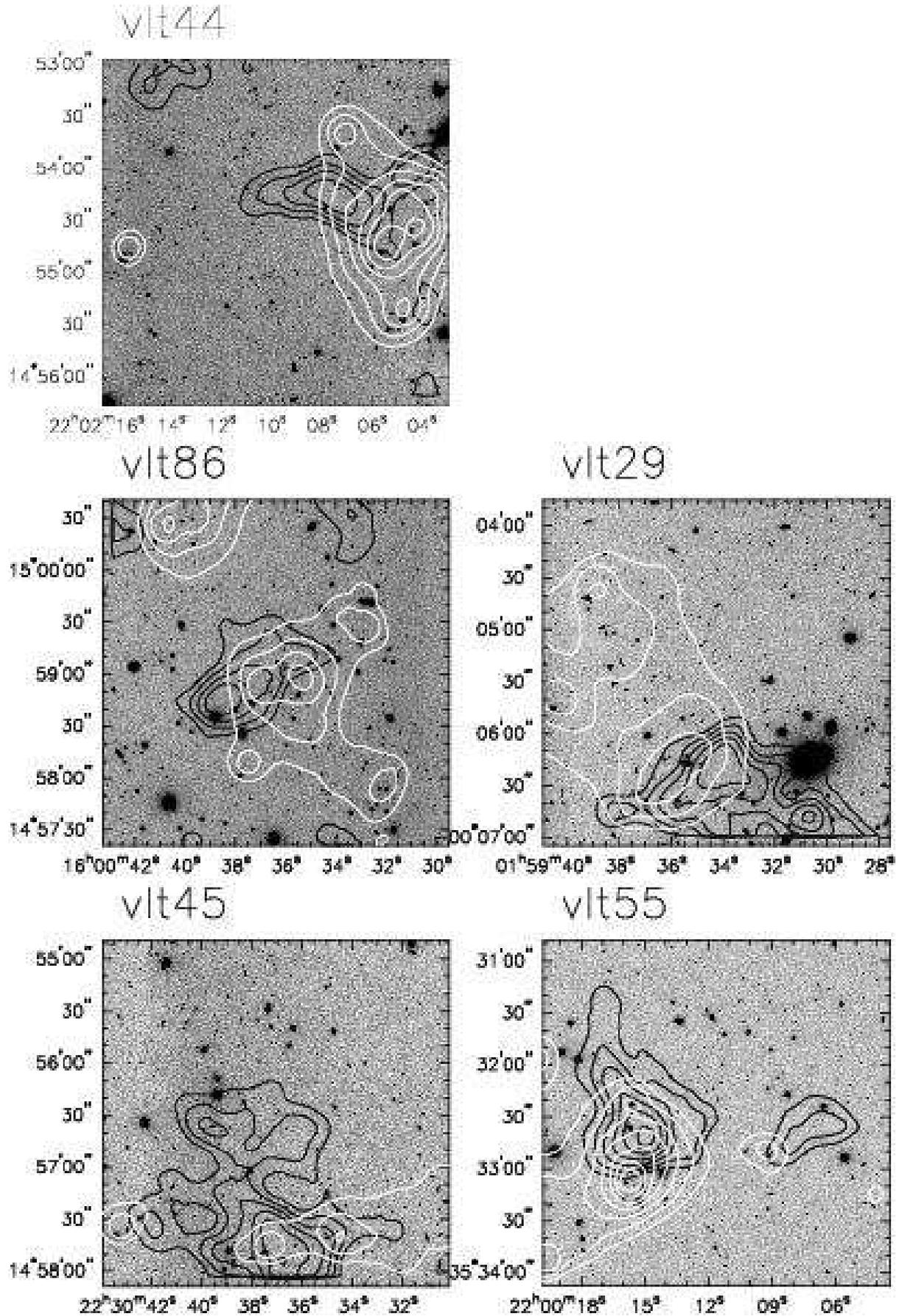}
\caption{$3^\prime\times 3^\prime$ clip of the most promising cluster candidates. 
The $M_{\rm ap}$-contours are black ($snr$-values are 1.5, 2.0, 2.5, 3.0, 3.5, 4.0) and the light distribution contours are white ($snr$-values are 1, 2, 3, 4, 5).}
\label{fig:cluster_can}
\end{figure*}

We performed an analysis of 50 VLT fields using the halo-filter function and obtained weak lensing maps of the fields.
In 12 of these maps the presence of significant $M_{\rm ap}$-peaks is revealed that are associated with 
overdensities in the light distribution (Fig. \ref{fig:cluster_candidateI} and Fig. \ref{fig:cluster_candidateII}).
Five of these remain after a careful analysis (we calculated the light distribution, used different source catalogues, analysed the fields with various filter radii, excluded high ellipticities and calculated the tangential shear profile).
In Fig. \ref{fig:cluster_can} $3^\prime\times 3^\prime$ clips of the five most promising cluster candidates (vlt44, vlt86, vlt29, vlt45, vlt55) are displayed. 
One of these candidates has been discovered before by \citet{mvm01} (candidate vlt55); see ESO Press Release 24/00.
From the simulations we expect to find about two clusters with a $snr$ larger than four using a fixed filter radius of $\theta_0=3\myarcmin 8$.
This is comparable with the analysed data, as one candidate (vlt55) has a $snr$ larger than four using the halo-filter with $\theta_0=3\myarcmin 8$ and the $snr$ of two further candidates (vlt29 and vlt45) reaches four for a slightly different filter radius.

The current data at hand (one colour; $I$-band; small field-of-view)
allowed us to detect these mass concentrations, but do not
permit a more detailed analysis at this stage. As our candidates are
not in the centre of the FORS1-fields, the gravitational shear and hence
the mass distribution of the clusters cannot reliably be mapped over 
a large range of radii. With one colour, we cannot obtain
an estimate for the cluster redshifts or for the redshift distribution
of the galaxy population used for the analysis. This, besides an
elimination of the contribution from foreground galaxies, is essential
to obtain accurate mass estimates.
Hence, to make further progress with these candidates we require multi-colour observations and a larger field-of-view.

\begin{table*}
\caption{List of all 50 VLT fields. $n5$ and $n3$ are the number density of galaxies per square arcminute in each field of the final catalogue with SExtractor parameter settings $N=5$ and $N=3$ connected pixels with $k=2\sigma$ and $k=1\sigma$ above the mean background value, respectively. 
The seeing given in this table correspond to the stacked image. $N(>3)$ denotes the number of peaks per field with a larger $snr$ than $3$.
For the cluster candidates the maximum $snr$ and the corresponding halo-filter radius are listed. 
The maximum $snr$ of the rejected candidates are in brackets. 
}
\label{SumTab}
\begin{center}
\begin{tabular}{l|c|r|c|r|r|r|r|r}
name &RA (J2000) & DEC (J2000) & seeing & $n5$ & $n3$ & $I_{AB}$ [mag] & $N(>3)$ & $snr_{\rm max}(\theta_0)$ \\
&&&&&&&&\\
\hline
vlt27 &00 59 28.1 &$-$00 18 28 &$0\myarcsec 72$ &  8 & 15 & 24.5 & 3 &  \\
vlt28 &01 31 40.3 &$-$00 22 28 &$0\myarcsec 54$ & 27 & 40 & 24.9 & 0 &  \\
vlt29 &01 59 40.8 &$-$00 03 51 &$0\myarcsec 49$ & 32 & 36 & 25.1 & 2 & 4.4 ($5\myarcmin 5$) \\
vlt30 &02 28 44.0 &$-$00 03 26 &$0\myarcsec 54$ & 25 & 38 & 24.9 & 3 &  \\
vlt31 &01 00 21.8 &$-$03 15 31 &$0\myarcsec 57$ & 24 & 35 & 25.0 & 5 &  \\
vlt33 &02 00 08.1 &$-$03 00 30 &$0\myarcsec 44$ & 29 & 44 & 24.9 & 5 &  \\
vlt35 &00 59 35.3 &$-$06 10 05 &$0\myarcsec 73$ & 17 & 24 & 24.8 & 3 &  \\
vlt36 &01 28 53.1 &$-$06 01 39 &$0\myarcsec 68$ & 25 & 26 & 24.8 & 3 & [4.3 ($1\myarcmin 4$)]  \\
vlt37 &01 57 05.8 &$-$06 05 01 &$0\myarcsec 90$ & 14 & 18 & 24.5 & 1 &  \\
vlt39 &21 30 45.3 &$-$09 58 45 &$0\myarcsec 76$ & 18 & 23 & 24.8 & 1 &  \\
vlt40 &22 04 37.9 &$-$10 15 09 &$0\myarcsec 71$ & 19 & 27 & 24.9 & 3 &  \\
vlt42 &22 29 29.2 &$-$10 12 01 &$0\myarcsec 72$ & 18 & 17 & 24.7 & 2 & [3.8 ($3\myarcmin 8$)]  \\
vlt43 &21 30 25.3 &$-$15 11 48 &$0\myarcsec 55$ & 20 & 29 & 25.0 & 4 &  \\
vlt44 &22 02 16.6 &$-$14 53 03 &$0\myarcsec 64$ & 24 & 25 & 24.9 & 2 & 4.4 ($1\myarcmin 4$)  \\
vlt45 &22 30 41.8 &$-$14 54 55 &$0\myarcsec 46$ & 28 & 32 & 24.8 & 5 & 4.1 ($3\myarcmin 4$) \\
vlt46 &22 01 42.2 &$-$20 10 55 &$0\myarcsec 65$ & 24 & 33 & 25.0 & 4 &  \\
vlt47 &22 29 33.8 &$-$20 14 44 &$0\myarcsec 51$ & 25 & 38 & 25.0 & 1 &  \\
vlt48 &21 30 45.3 &$-$24 53 40 &$0\myarcsec 63$ & 22 & 30 & 24.8 & 1 &  \\
vlt49 &21 58 44.7 &$-$24 57 15 &$0\myarcsec 62$ & 21 & 30 & 24.6 & 3 & \\
vlt50 &22 30 43.8 &$-$25 01 42 &$0\myarcsec 55$ & 24 & 36 & 25.0 & 1 &  \\
vlt51 &20 59 30.5 &$-$30 18 31 &$0\myarcsec 62$ & 15 & 22 & 24.5 & 1 &  \\
vlt52 &22 00 26.2 &$-$30 01 45 &$0\myarcsec 65$ & 19 & 26 & 24.8 & 1 &  \\
vlt53 &22 31 15.3 &$-$30 07 15 &$0\myarcsec 55$ & 27 & 39 & 24.7 & 0 &  \\
vlt54 &21 29 53.6 &$-$34 51 52 &$0\myarcsec 57$ & 22 & 24 & 25.1 & 5 & [3.7 ($2^\prime$)]  \\
vlt55 &22 00 14.1 &$-$35 30 54 &$0\myarcsec 53$ & 22 & 33 & 24.6 & 4 & 4.3 ($3\myarcmin 8$)  \\
vlt56 &22 30 06.4 &$-$35 10 33 &$0\myarcsec 83$ & 13 & 19 & 24.5 & 1 &  \\
vlt57 &21 28 04.9 &$-$39 49 02 &$0\myarcsec 55$ & 24 & 33 & 25.0 & 3 &  \\
vlt58 &22 00 06.7 &$-$40 04 55 &$0\myarcsec 49$ & 26 & 37 & 24.8 & 1 &  \\
vlt59 &22 29 11.8 &$-$39 36 28 &$0\myarcsec 70$ & 21 & 27 & 24.7 & 1 &  \\
vlt60 &22 59 24.4 &$-$10 01 29 &$0\myarcsec 47$ & 28 & 33 & 25.0 & 4 & [4.1 ($1\myarcmin 4$)]  \\
vlt61 &22 59 24.2 &$-$15 08 47 &$0\myarcsec 47$ & 34 & 47 & 25.0 & 1 &  \\
vlt62 &22 59 01.8 &$-$19 44 03 &$0\myarcsec 47$ & 28 & 40 & 25.0 & 1 &  \\
vlt63 &22 59 39.5 &$-$24 52 51 &$0\myarcsec 49$ & 23 & 35 & 25.0 & 5 &  \\
vlt64 &22 59 56.1 &$-$30 14 27 &$0\myarcsec 60$ & 19 & 27 & 24.5 & 2 &  \\
vlt65 &23 00 44.3 &$-$34 55 26 &$0\myarcsec 54$ & 25 & 38 & 25.0 & 4 &  \\
vlt66 &23 01 24.8 &$-$40 25 20 &$0\myarcsec 77$ & 12 & 17 & 23.6 & 2 &  \\
vlt75 &21 28 14.7 &$-$20 07 18 &$0\myarcsec 56$ & 26 & 36 & 24.9 & 4 &  \\
vlt76 &21 32 21.0 &$-$30 25 57 &$0\myarcsec 63$ & 19 & 26 & 24.6 & 3 &  \\
vlt77 &14 59 07.4 & 00 07 54 &$0\myarcsec 80$ & 16 & 16 & 24.8 & 2 & [3.6 ($3\myarcmin 4$)]  \\
vlt78 &14 59 03.2 & 05 11 32 &$0\myarcsec 50$ & 31 & 34 & 25.0 & 1 & [3.2 ($5^\prime$)]  \\
vlt79 &14 59 32.7 & 10 13 19 &$0\myarcsec 76$ & 13 & 15 & 25.0 & 3 & [4.5 ($5^\prime$)]  \\
vlt80 &15 30 17.5 & 00 10 58 &$0\myarcsec 60$ & 23 & 32 & 24.8 & 3 &  \\
vlt81 &15 29 40.4 & 04 54 10 &$0\myarcsec 63$ & 20 & 27 & 24.7 & 1 &  \\
vlt82 &15 28 59.7 & 10 14 59 &$0\myarcsec 59$ & 22 & 31 & 24.6 & 4 &  \\
vlt83 &15 59 00.7 &$-$00 07 14 &$0\myarcsec 87$ & 11 & 16 & 24.3 & 4 &  \\
vlt84 &16 03 35.0 & 05 10 46 &$0\myarcsec 91$ & 11 & 15 & 23.8 & 0 &  \\
vlt85 &15 56 47.6 & 10 17 28 &$0\myarcsec 66$ & 19 & 27 & 24.7 & 0 &  \\
vlt86 &16 00 30.1 & 14 58 35 &$0\myarcsec 78$ & 16 & 16 & 24.6 & 1 & 4.0 ($1\myarcmin 7$) \\
\end{tabular}
\end{center}
\end{table*}
%
%
\section{Summary and conclusions}
\label{sec:conclusion}
We have analysed the ability of the $M_{\rm ap}$-statistics to detect massive mass concentrations.
We first maximised the effectiveness of the $M_{\rm ap}$-statistics using analytic descriptions and then applied $M_{\rm ap}$ to synthetic images created from numerical simulations.
We investigated the influence of image treatment and border effects on the $snr$ and number density of halos.
Finally, we applied $M_{\rm ap}$ to a data set obtained with the VLT and compared the results with the predictions obtained from the simulations and performed a detailed analysis of the cluster candidates.
Our major findings are as follows.
 
We created twelve synthetic images, each covering a $30^\prime\times 30^\prime$ area, from $N$-body simulations.
We compared the halo-filter and the polynomial filter and found that 4.5 times more peaks with a $snr$ larger than four are detected with the halo-filter compared to the polynomial filter. 
However, if we take into account the contamination ratio of noise peaks to the total number of peaks in the $M_{\rm ap}$-maps, the difference in efficiency to find real clusters for the two filter types is much less distinct.

We studied the effect of weighting and image treatment on the $snr$ of peaks in the weak lensing map.
We found that, on the one hand, the image treatment lowers the $snr$ significantly so that the expected number density of halos decreases by a factor of two.
On the other hand, weighting has only a weak influence on the $snr$.
To compare real data with simulations, both effects have to be taken into account.
Based on our findings in Sect. \ref{sec:imagetreat} we concluded that it is more efficient for future weak lensing surveys to propose for medium deep images than for a few very deep images.

We studied the influence of the border effects on the number density of peaks in the weak lensing maps.
For that, we subdivided the fields of the numerical simulations into 300 subfields (all having the same size as the VLT fields).  
We found that the border effects affect the $snr$ of clusters significantly and consequently has a large impact on the number density of peaks in the $M_{\rm ap}$-maps.
The number density of peaks in the $M_{\rm ap}$-maps decreases by a factor of two.
Including the border effects, we expect for VLT-sized images $\sim 3$ peaks per square degree with a $snr$ larger than four using the halo filter with a filter radius of $\theta_0=3\myarcmin 7$, see Table \ref{tab:numberdensity}.
In the future, ground-based wide-field images will be common, so this effect will not play an important role.
However, for future space-based missions with smaller field of views (compared to ground-based cameras) this effect still has to be taken into account.

We then performed a statistical peak analysis of the 50 VLT fields and found that the number density of peaks (for $snr > 3$ and a filter scale of $3\myarcmin 8$) is comparable with the number density obtained from the numerical simulations, taking into account border effects and image treatment.
However, for a filter scale of $5\myarcmin 7$ the number density of peaks in the VLT fields is significantly smaller compared to the simulations and is comparable with the number density of noise peaks, see Fig. \ref{fig:sim-VLT}.
We point out again that the reason for this could be that low-mass clusters are present in the VLT fields which are not matched with the larger filter function.
If this would be the case, then this is a way to constrain the dark matter halo size in a statistical way.


Finally, we reported the results of a detailed analysis of the VLT fields.
We detected several $M_{\rm ap}$-peaks with a $snr > 3.0$ in our 50 fields.
For 12 of the $M_{\rm ap}$-peaks we could associate an overdensity in the light distribution. 
These cluster candidates were analysed in detail by using different filter radii, SExtractor parameter settings and ellipticity cuts.
Furthermore, the tangential ellipticity and its dependence on distance to the $snr$-maximum was analysed.
Finally, five promising candidates remain after selection which need a follow-up observation in different filters to clarify their nature. 

In this work we are quite sceptical on the efficiency of finding {\it individual} clusters with masses less then $M\approx 3\times 10^{14}\,M_{\sun}$ for redshifts larger than $z=0.3$.
But since there is much more information in low-mass clusters than in high mass ones, simply because they are so much outnumbered,
the goal should not be a blind search for {\it individual} clusters, but a {\it statistical} blind search, especially for future large weak lensing surveys.
A statistical peak analysis has already been applied to observable data by \citet{mhs02} and is a valuable tool to explore cosmological models. 
The major part of our work should be seen as a foundation stone for the comparison of weak lensing surveys with ray-tracing through $N$-body simulations since we now consider image treatment or other observational effects (like the border effect). 
This will improve the manner by which cosmological models can be explored with galaxy clusters.
With better synthetic data (more realistic background galaxies could be simulated with shapelets, for instance) generated from ray-tracing through $N$-body simulations for different cosmological models and a much larger survey (compared to our 0.64 square degree) it would then be possible to constrain cosmological parameters and especially the statistics about the distribution of dark matter.     

%
%
\begin{appendix}
\section{Calculation of the number density of halos}
\label{appendixa}
We first describe in this appendix in detail the calculations to obtain the aperture mass, $M_{\rm ap}$, given a truncated NFW-profile, a redshift distribution of background galaxies and the new filter function (halo-filter).
We show then how we calculate the expected number density of halos assuming a halo distribution which utilise the Press-Schechter theory.
The calculations are restricted to a flat universe ($\Omega_{\rm m}+\Omega_\Lambda=1$).

The aperture mass, $M_{\rm ap}$, is given by
\be
   M_{\rm ap}=\int {\rm d}^2\mytheta\int_{z_{\rm l}}^{\infty} \dd z_{\rm s}\,\kappa(\mytheta, z_{\rm l}, z_{\rm s})\,U(\vartheta)\,p(z_{\rm s}),
\label{zzmap}
\ee
%
%
where $U$ is the halo-filter obtained by using Eq. \ref{relation} and Eq. \ref{filter}, $p$ is the redshift distribution of galaxies given in Eq. \ref{redshiftdistribution}.

The convergence, $\kappa$, is that of an NFW-profile and is truncated at the virial radius, $r_{\rm vir}$, see \citet{taj03} and is given by
\be
\kappa(y)=3\Omega_{\rm m}\delta_{\rm s}r_{\rm s}\left(\frac{H_0}{c}\right)^2\frac{D_{\rm l}D_{\rm ls}}{D_{\rm s}}f(y),
\ee
with $y=r/r_{\rm s}$, where $r_{\rm s}$ is the scale radius, $D$ denotes the angular diameter distance and the function $f$ is given by,
\begin{displaymath}
f(y)=\left\{
\begin{array}{lr}
A+(1-y^2)^{-3/2}\,{\rm arccosh} \frac{y^2+c_{\rm N}}{y(1+c_{\rm N})}; & y<1\\
\frac{c^2_{\rm N}-1}{3(1+c_{\rm N})}\frac{2+c_{\rm N}}{1+c_{\rm N}}; & y=1\\
A-(y^2-1)^{-3/2}\,\arccos \frac{y^2+c_{\rm N}}{y(1+c_{\rm N})}; & 1<y\leq c_{\rm N}\\
0; & y>c_{\rm N}
\end{array}
\right.
\end{displaymath}
where 
\be
A=\frac{y^2+c_{\rm N}}{y(1+c_{\rm N})}.
\ee
The quantity $\delta_{\rm s}$ is given by
\be
\delta_{\rm s}=\frac{\delta_{\rm vir}}{3}\frac{c_{\rm N}^3}{\log(1+c_{\rm N})-c_{\rm N}/(1+c_{\rm N})}.
\ee
The concentration parameter of an NFW profile is
\be
c_{\rm N}=\frac{r_{\rm vir}}{r_{\rm s}}
\ee
and can also be expressed by \citep{bks01}
\be
c_{\rm N}=\frac{c_*}{1+z}\left(\frac{M}{10^{14}h^{-1}M_{\sun}}\right)^{-0.13},
\label{concentration}
\ee
where we set $c_*=8$ for an open universe.
The mass within a sphere of radius $r_{\rm vir}$ (virial radius) is 
\be
M_{\rm vir}=\frac{4\pi\rho_{\rm s}r^3_{\rm vir}}{c^3_{\rm N}}\left[\log(1+c_{\rm N})-\frac{c_{\rm N}}{1+c_{\rm N}}\right].
\ee
The virial mass can also be defined by the spherical top-hat collapse model as
\be
M_{\rm vir}=\frac{4\pi}{3}\delta_{\rm vir}(z)\bar\rho_0 r^3_{\rm vir},
\ee
with
\be
\bar\rho_0=\rho_{\rm crit}\Omega_{\rm m},
\ee
where $\rho_{\rm crit}$ is the critical density.
The virial overdensity reads \citep{bks01}
\be
\delta_{\rm vir}(z)=(18\pi^2+82x-39x^2)\frac{1}{\Omega(z)},
\ee
where $x\equiv\Omega(z)-1$ and
\be
\Omega(z)=\frac{(1+z_{\rm l})^3\Omega_{\rm m}}{(1+z_{\rm l})^3\Omega_{\rm m}+(1+z_{\rm l})^2(1-\Omega_{\rm m}-\Omega_\Lambda)+\Omega_\Lambda}.
\ee

To calculate the expected number density of halos per steradian with aperture mass larger than $M_{\rm ap}$ we use the formula derived by \citet{krs99},
\be
N(>M_{\rm ap})=\frac{c}{H_0}\int\dd z_{\rm l}\frac{(1+z_{\rm l})^2}{E(z_{\rm l})}D_{\rm l}^2(z_{\rm l})\tilde G(z_{\rm l},M_{\rm ap}),
\ee
with
\be
\tilde G(z_{\rm l},M_{\rm ap})=\int_{M_{\rm t}(M_{\rm ap},z_{\rm l},\theta)}^\infty \dd M\,N_{\rm halo}(M,z_{\rm l})
\ee
and 
\be
E(z_{\rm l})\!=\!\sqrt{(1\!+\!z_{\rm l})^3\Omega_{\rm m}\!+\!(1\!+\!z_{\rm l})^2(1-\!\Omega_{\rm m}-\!\Omega_\Lambda)\!+\Omega_\Lambda}
\ee
and
\begin{eqnarray}
N_{\rm halo}(M,z_{\rm l})\,\dd M \dd V_{\rm c} = & \sqrt{\frac{2}{\pi}}\,\frac{\bar \rho}{M}\frac{\delta_{\rm crit}(z_{\rm l})}{\sigma^2(M)}\,\left|\frac{\dd \sigma(M)}{\dd M}\right|\nonumber\\
\times & \exp\left(-\frac{\delta^2_{\rm crit}(z_{\rm l})}{2\sigma^2(M)}\right) \dd M \dd V_{\rm c} 
\end{eqnarray}
is number of objects in the comoving volume $\dd V_{\rm c}$ with mass in the interval $\dd M$.
The quantity
\be
\delta_{\rm crit}(z_{\rm l})=\frac{\delta^0_{\rm crit}}{D_+(z_{\rm l},\Omega_0,\Omega_\Lambda)}
\ee
is the critical density threshold for spherical collapse which depends on the linear growth factor, $D_+$ \citep{lac93}.
The quantity $\sigma(M)$ is the present linear theory rms density fluctuation with shape parameter $\Gamma$ and normalisation $\sigma_8$.
As \citet{krs99} we use the fitting formulae A6-A19 given in the appendix of \citet{nfw97} to calculate $\sigma(M)$ and $\delta^0_{\rm crit}$.

\end{appendix}

\begin{acknowledgements}
We would like to thank Peter Watts for helpful discussions, 
Oliver Czoske and Nadja M\"uller for their careful reading of the manuscript.
Furthermore, we thank Takashi Hamana for very kindly allowing us to use his simulations, and
the service observing team at Paranal for providing us with this excellent data set. 
Also we would like to thank the anonymous referee for his rapid examination of our paper and for his 
knowledgeable suggestions.
This work was supported 
by the German Ministry for Science and Education (BMBF) through DESY
under the project 05AE2PDA/8,
and by the Deutsche Forschungsgemeinschaft under the project
SCHN 342/3--1.
Roberto Maoli thanks the City of Paris for funding his research grants,
at IAP and the IAP institute for local supports.
\end{acknowledgements}

\end{document}